\pdfoutput=1
\documentclass[11pt]{article}

\usepackage{acl}

\usepackage{times}
\usepackage{latexsym}
\usepackage{amsmath}
 \usepackage{multirow}
 \usepackage{booktabs}
\usepackage[T1]{fontenc}
\usepackage[utf8]{inputenc}

\usepackage{microtype}

\usepackage{inconsolata}

\usepackage{graphicx}
\usepackage{float}
\usepackage{placeins,enumitem}

\title{SVeritas: Benchmark for Robust Speaker Verification under Diverse Conditions}

\author{
 \textbf{Massa Baali\textsuperscript{1}},
 \textbf{Sarthak Bisht\textsuperscript{1}},
 \textbf{Francisco Teixeira
\textsuperscript{2}},
 \textbf{Kateryna Shapovalenko \textsuperscript{1}},
\\
 \textbf{Rita Singh \textsuperscript{1}},
 \textbf{Bhiksha Raj \textsuperscript{1}},
\\
 \textsuperscript{1}  Carnegie Mellon University, Pittsburgh, USA \\
 \textsuperscript{2}INESC-ID, Lisbon, Portugal
\\
 \texttt{mbaali@cs.cmu.edu}
}

\begin{document}
\maketitle
\begin{abstract}
Speaker verification (SV) models are increasingly integrated into security, personalization, and access control systems, yet their robustness to many real-world challenges remains inadequately benchmarked. %Real-world systems can face diverse conditions, some naturally occurring, and others that may be purposely, or even maliciously created, which introduce mismatches between enrollment and test data, affecting their performance. 
These include a variety of natural and maliciously created conditions causing signal degradations or mismatches between enrollment and test data, impacting performance.
%Ideally, the effect of all of these on model performance must be benchmarked; however existing benchmarks fall short, generally evaluating only a subset of potential conditions, and missing others entirely. 
Existing benchmarks evaluate only subsets of these conditions, missing others entirely.
%We introduce \textit{SVeritas}, the \textbf{S}peaker \textbf{Veri}fication \textbf{tas}ks benchmark suite, which evaluates the performance of speaker verification systems under an extensive variety of stressors, including ``natural'' variations such as duration, spontaneity and content of the recordings, background conditions such as  noise, microphone distance, reverberation, and channel mismatches, recording condition influences such as audio bandwidth and the effect of various codecs, physical influences, such as the age and health conditions of the speaker, as well as the suspectibility of the models to spoofing and adversarial attacks.
We introduce SVeritas, a comprehensive Speaker Verification tasks benchmark suite, assessing SV systems under stressors like recording duration, spontaneity, content, noise, microphone distance, reverberation, channel mismatches, audio bandwidth, codecs, speaker age, and susceptibility to spoofing and adversarial attacks.
While several benchmarks do exist that each cover some of these issues, SVeritas is the first comprehensive evaluation that not only includes all of these, but also several other entirely new, but nonetheless important real-life conditions that have not previously been benchmarked. We use SVeritas to evaluate several state-of-the-art SV models and observe that while some architectures maintain stability under common distortions, they suffer substantial performance degradation in scenarios involving cross-language trials, age mismatches, and codec-induced compression. Extending our analysis across demographic subgroups, we further identify disparities in robustness across age groups, gender, and linguistic backgrounds. By standardizing evaluation under realistic and synthetic stress conditions, SVeritas enables precise diagnosis of model weaknesses and establishes a foundation for advancing equitable and reliable speaker verification systems.

% Read vs spontaneous speech
% - Speaking same vs different sentence
% - Speaking same vs different language
% - Same vs different channel (codec/bandwidth)
% - Far vs near field (distance from microphone)
% - Real vs synthetic speech (spoofing)
% - Comparing different synthetic speakers
% - Performance under various noise conditions
% - Performance under adversarial attacks
% - duration
% - age

% of covering essential real-world and synthetic factors, such as channel mismatch, cross-lingual variability, demographic shifts, background noise, reverberation, speech codecs, bandwidth limitations, duration variability, and adversarial attacks. We introduce SVeritas, the Speaker VERification TASks benchmark comprising diverse evaluation scenarios that systematically span these dimensions. We use SVeritas to evaluate the robustness of several state-of-the-art SV models and observe that while some architectures maintain stability under common distortions, they suffer substantial performance degradation in scenarios involving cross-language trials, age mismatches, and codec-induced compression. Extending our analysis across demographic subgroups, we further identify disparities in robustness across age groups, gender, and linguistic backgrounds. By standardizing evaluation under realistic and synthetic stress conditions, SVeritas enables precise diagnosis of model weaknesses and establishes a foundation for advancing equitable and reliable speaker verification systems.
\end{abstract}

\section{Introduction}\label{sec:intro}
Speaker verification technology has achieved remarkable accuracy under controlled conditions, driven by advances in deep neural embeddings, margin-based losses, and self-supervised pretraining. However, real-world deployments -- from secure access control and telephony authentication, to personalized assistants, and law enforcement -- confront a broad spectrum of challenges that degrade performance, including degradations to the signal itself, and mismatches between the test utterances and the enrollment recordings they are compared to. These mismatches arise from natural variability (e.g., spontaneous versus read speech, cross-language trials, or temporal drift), environmental distortions (e.g., reverberation, background noise, far- versus near-field capture), device and codec artifacts, demographic factors (age, health or physical condition), and even malicious manipulations such as spoofing or adversarial attacks. Without comprehensive, standardized evaluation across these diverse stressors, it remains unclear which aspects of SV systems are robust in practice and where critical vulnerabilities lie.

Existing benchmarks each target a narrow subset of these challenges. For example, CommonBench \cite{hintzcommonbench} offers large-scale multilingual text-independent trials, yet it relies on an ECAPA-based outlier filter that may prune precisely the hardest cases (e.g., heavy accents or noisy recordings), and omits deliberate distortions such as codec compression or spoofed audio. IndicSUPERB \cite{javed2023indicsuperb} highlights performance on twelve Indian languages but focuses exclusively on scripted, read speech in clean or synthetic noise conditions, neglecting cross-language scenarios, far-field capture, or adversarial manipulations. Other benchmarks examine specific dimensions in isolation -- far-field effects in MultiSV \cite{movsner2022multisv}, age variation in time-varying SV \cite{doddington2012effect}, or spoofing attacks in ASVspoof \cite{wu2017asvspoof}. To the best of our knowledge, no prior suite spans the full gamut of natural, environmental, demographic, codec, and adversarial factors under a unified framework. Furthermore, many datasets lack sufficient metadata to analyze fairness across age, gender, or linguistic subgroups, rely on relatively large enrollment durations, or assume static, text-independent protocols that do not reflect modern low-resource, multi-file, or cross-domain requirements.

To address these gaps, we introduce SVeritas, the first comprehensive speaker verification benchmark suite that systematically evaluates state-of-the-art models across an extensive set of real-world and synthetic stressors. SVeritas assembles trials spanning (i) content and style variations (read vs. spontaneous, same vs. different sentences, multi-language), (ii) acoustic and channel mismatches (noise types and levels, far- vs. near-field, codec and bandwidth variations), (iii) demographic and physical factors (age-group mismatches, health or emotional states), (iv) enrollment/test duration and multi-file enrollment, (v) security threats (spoofing via TTS/VC pipelines, universal and adversarial perturbations), and (vi) speaking-style adaptation (Lombard speech under noise-induced conditions). By unifying these dimensions, SVeritas not only measures aggregate metrics such as equal error rate and detection cost function, but also facilitates fine-grained analyses of performance disparities across demographic subgroups and operating conditions. Through extensive evaluation of leading architectures, %(e.g., \textbf{<FINAL MODEL LIST>})
 we uncover systemic weaknesses -- particularly in cross-language, age-mismatch, and codec-compressed trials -- and expose fairness gaps that vary nontrivially by gender and language background.

In summary, SVeritas establishes a rigorous, reproducible foundation for diagnosing robustness and equity in speaker verification. By revealing hitherto uncharacterized vulnerabilities and enabling targeted stress-testing, our benchmark paves the way for developing more reliable, inclusive, and secure SV systems suitable for deployment in the complex acoustic and demographic landscapes of real-world applications. Our code is publicly available with documentation, fostering straightforward reproducibility and extensibility. \url{https://github.com/massabaali7/SVeritas}.

\begin{figure*}[t]
  \centering
  \includegraphics[width=\linewidth]{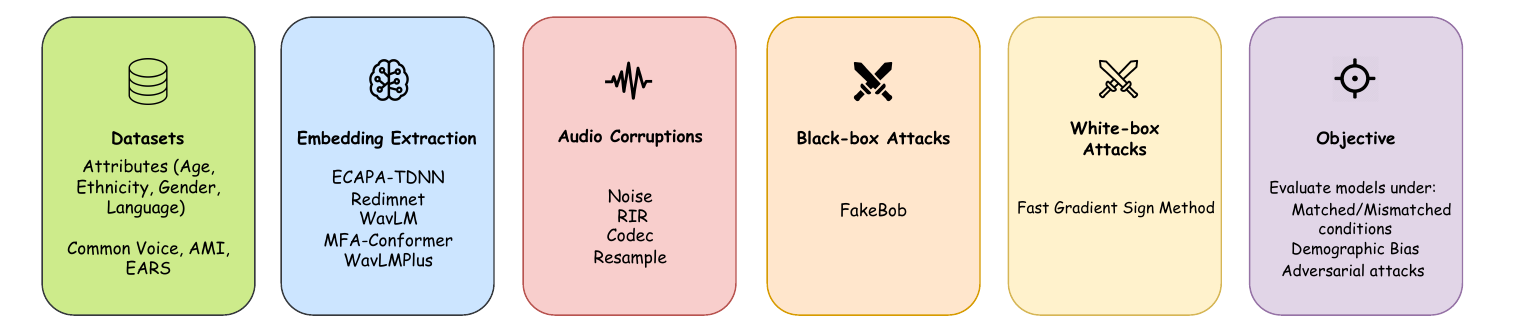}
  \caption{Overview of our benchmark \textsc{SVeritas}.}
  \label{fig:sveritasFig}
\end{figure*}

\section{Background and Related Work}
\subsection{Speaker Verification Systems and their Vulnerabilities}
Speaker Verification systems attempt to verify the identity of a speaker by comparing (embeddings) derived from) their voice recordings to a template such as a statistical model \cite{reynolds2000speaker}, or other embeddings derived from ``enrollment'' data \cite{dehak2010front}. 
%The latter approach has generally been more successful and is the format employed in most SOTA systems, with the key focus being on how the embeddings are derived from the data [CITES] and the actual data that the models deriving these embeddings are trained from [CITES].

Besides the models themselves, the performance of the SV system depends on many other factors, primary being the native quality of the signal itself. The best performances are generally obtained with studio-quality broadband signals \cite{villalba2020state}, which degrades when the bandwidth of the signal is restricted, such as over telephony or cell-phone channels \cite{kenny2010bayesian}. The application of various codecs also degrades performance \cite{njegovec2025forensic}. External influences such as background noise, recording room responses and reverberation also degrade performance \cite{ko2017study} \cite{nandwana2018robust}. Perhaps most concerningly, innate biases within the system too result in reduced performances for some categories of subjects \cite{hajavi2023study}. Performance is also dependent on the duration of the recording (longer recordings are better \cite{poddar2018speaker}), and on whether the speech is spontaneous or recited, e.g. by reading \cite{nakamura2008differences}.

A second, and equally important source of degradation is \textit{mismatches} between the conditions of the test and enrollment data.  Signal differences in bandwidth, channel condition, duration \textit{etc.} can result in degraded performance. \textit{Content} variations, such as language and dialectal differences \cite{abdullah2025dialect}, as well as exactly what is spoken \cite{dey2018phonetic} can cause degradations. \textit{Biological} influences, such as changes in the age or health status of the speaker too can cause degradations \cite{kelly2011effects}.

A third and increasingly important source of degradation is \textit{active} misdirection, such as through voice mimicry \cite{hautamaki2013vectors}, synthetic voice recordings \cite{zuo2024advtts} or adversarial modifications \cite{alzantot2018did}\cite{zhou2023adversarial}\cite{jati2021adversarial} which can make an SV system fraudulently accept an imposter or reject a genuine match.

\subsection{Remediations}
The most common approach to remediation of natural and mismatch-based degradations is through inclusive training -- adding data with the variations that one must be robust to in the training data of the model \cite{ko2017study}, \textit{e.g.}
%spontaneity is achieved by including spontaneous speech in the training [CITE], 
far-field and noisy conditions (often through simulated room responses and digital addition of noise) \cite{ko2017study, yakovlev2024reshape, thienpondt2023ecapa2, al2021mitigate},  codec and bandwidth variations \cite{polacky2016impact}, \textit{etc.}. Explicit modeling of, and compensation for effects such as noise and reverberation has also been found to be effective in some settings \cite{al2021mitigate}.
%[DENOISING].

Another popular approach to mitigate the effect of variations is through contrastive losses that attempt to neutralize variations by minimizing the distances of (embeddings from) recordings and their mismatched counterparts \cite{inoue2020semi}. Alternate methods attempt to disentangle confounding sources of variation \cite{nam2024disentangled}.

Defenses against misdirection attacks include explicit attempts at detecting mimicry \cite{hautamaki2013vectors}, or through adversarial defenses such as adversarial training, which can protect to some extent against adversarial attacks \cite{zhou2023adversarial}.

\subsection{Benchmarking}
Speaker verification systems are often used in critical settings, such as user authentication or law enforcement. Consequently, benchmarking their performance under these various challenges becomes necessary.

Indeed, benchmarking has been central to the development of SV systems, guiding progress through standardized evaluation protocols. Traditional efforts such as the NIST Speaker Recognition Evaluations (SREs) \cite{sadjadi20222021,sadjadi20172016} have driven advances in SV for over two decades, though their design primarily targets constrained settings involving telephone and microphone speech. The VoxCeleb Speaker Recognition Challenge (VoxSRC) \cite{nagrani2020voxsrc} was introduced to evaluate the ability of modern speaker recognition systems to identify speakers from speech captured ‘in the wild’, %providing a large-scale, unconstrained benchmark built from audio sourced from online videos. However, these benchmarks still fall short in systematically testing robustness to specific distortions and real-world variability. 
To address isolated robustness factors, several specialized benchmarks have been introduced. The Short-Duration Speaker Verification Challenge (SDSVC) \cite{zeinali2019short} and Far-Field SVC \cite{qin2020interspeech} focus on duration and spatial variability. SUPERB \cite{yang2021superb} offers a comprehensive suite for evaluating speech representation learning across multiple tasks, including speaker verification, but the SV component remains relatively coarse-grained and lacks detailed stress testing.
% % 
More recently, VoxBlink \cite{lin2024voxblink} emphasized robustness to device mismatch and short-duration utterances, uncovering substantial performance degradation under realistic deployment scenarios. Other efforts at benchmarking have been mentioned in Section \ref{sec:intro}. 

Notably, while each of these benchmarks evaluates the system under subsets of the various challenges a real-life deployment may face, unlike other speech pattern recognition tasks, such as speech recognition \cite{shah2024speech}, there is as yet no single benchmark suite that integrates all of the broader robustness dimensions such as recording condition variations, demographic variation, adversarial perturbations, and codec-induced compression into a unified diagnostic framework. Some factors related to demographics and content are not addressed by any existing benchmark. SVeritas addresses this gap.

\section{SVeritas Benchmark}
\begin{table*}[t]
  \centering
  \resizebox{\textwidth}{!}{
  \begin{tabular}{lccccc}
    \hline
    \textbf{Condition} & \textbf{WavLM-Base} & \textbf{WavLM-Base+} & \textbf{RedimNet} & \textbf{ECAPA-TDNN} & \textbf{MFA-Conformer} \\
    \hline
    Real vs. Synthetic & 25.74\% & 23.76\% & 5.94\% & 5.94\% & 0.00\% \\
    FGSM         & 48.38\% & 48.39\% & 37.63\% & 52.63\% & 45.26\% \\
    FakeBob      & 25.81\% & 18.28\% & 10.75\% & 62.36\% & 35.48\% \\
    \hline
  \end{tabular}}
  \caption{\label{tab:merged_conditions}EARS: 
    EER for SV models under clean, spoofing, and adversarial attack conditions. 
  }
\end{table*}
SVeritas aims to provide a thorough benchmarking of SV systems, evaluating its performance under various degradations, mismatches, sources of bias, and attacks, providing both detailed and summarized evaluations, along with statistical significance reports where appropriate.  The tests are not only intended to evaluate the performance of the system under various conditions and threats that may be expected in real-life deployments, but to also provide a diagnostic tool to identify weaknesses, and detect any systematic biases or vulnerabilities. 
%While the results reported in this paper focus on summary analyses, the actual package provides a much more detailed and fine-grained analysis.
\begin{table*}[t]
\centering
\resizebox{\textwidth}{!}{
\begin{tabular}{llcccccc}
\hline
\textbf{Category} & \textbf{Subgroup} & \textbf{WavLM-Base} & \textbf{WavLM-Base+} & \textbf{RedimNet} & \textbf{ECAPA-TDNN} & \textbf{MFA-Conformer} & \textbf{Titanet} \\
\hline
\multirow{2}{*}{Gender} 
& Female (59 spks) & 13.74\% & 10.86\% & 1.50\% & 3.72\% & 2.44\% & 5.59\% \\
& Male (43 spks)   & 17.26\% & 12.77\% & 1.79\% & 4.41\% & 2.76\% & 5.71\% \\
\hline
\multirow{11}{*}{Age} 
& F (18–25), 13 spks & 13.01\% & 10.97\% & 2.34\% & 7.00\% & 4.62\% & 7.41\% \\
& F (26–35), 13 spks & 15.26\% & 12.71\% & 1.80\% & 4.24\% & 3.11\% & 6.93\% \\
& F (36–45), 7 spks  & 10.91\% & 8.30\%  & 0.27\% & 1.41\% & 1.07\% & 4.36\% \\
& F (46–55), 14 spks & 14.25\% & 11.99\% & 1.47\% & 3.31\% & 2.49\% & 5.38\% \\
& F (56–65), 10 spks & 16.84\% & 15.04\% & 1.52\% & 3.07\% & 1.83\% & 4.80\% \\
& F (66–75), 2 spks  & 26.28\% & 18.63\% & 0.73\% & 3.67\% & 1.61\% & 1.68\% \\
& M (18–25), 14 spks & 23.35\% & 16.85\% & 3.61\% & 7.81\% & 4.99\% & 10.52\% \\
& M (26–35), 10 spks & 16.16\% & 13.75\% & 2.02\% & 3.72\% & 2.78\% & 5.77\% \\
& M (36–45), 10 spks & 14.22\% & 10.79\% & 1.78\% & 3.43\% & 1.88\% & 6.02\% \\
& M (46–55), 4 spks  & 23.40\% & 18.07\% & 2.50\% & 7.89\% & 4.04\% & 7.20\% \\
& M (56–65), 5 spks  & 26.21\% & 19.52\% & 2.16\% & 6.43\% & 4.46\% & 11.35\% \\
\hline
\multirow{8}{*}{Ethnicity}
& F, White (40 spks)     & 14.67\% & 11.99\% & 1.44\% & 3.90\% & 2.44\% & 6.23\% \\
& F, Hispanic (4 spks)   & 8.12\%  & 5.88\%  & 0.43\% & 2.02\% & 1.24\% & 4.07\% \\
& F, Black (13 spks)     & 15.70\% & 13.63\% & 2.34\% & 5.18\% & 3.68\% & 6.66\% \\
& F, Asian (2 spks)      & 6.51\%  & 2.24\%  & 0.95\% & 6.59\% & 2.09\% & 5.34\% \\
& M, White (31 spks)     & 19.18\% & 14.30\% & 1.94\% & 4.87\% & 2.92\% & 7.21\% \\
& M, Hispanic (5 spks)   & 20.74\% & 15.86\% & 1.30\% & 5.97\% & 3.28\% & 8.99\% \\
& M, Black (5 spks)      & 16.23\% & 15.26\% & 1.47\% & 3.10\% & 1.50\% & 5.10\% \\
& M, Asian (2 spks)      & 17.58\% & 9.15\%  & 0.06\% & 0.30\% & 0.06\% & 2.45\% \\
\hline
\end{tabular}
}
\caption{\label{tab:demographic_eer}EARS: EER for SV models across gender, age, and ethnicity subgroups.}
\end{table*}
\begin{table*}[t]
\centering
\resizebox{\textwidth}{!}{
\begin{tabular}{llcccccc}
\hline
\textbf{Age}            & \textbf{Gender} & \textbf{WavLM-Base} & \textbf{WavLM-Base+} & \textbf{RedimNet} & \textbf{ECAPA} & \textbf{MFA-Conformer} & \textbf{Titanet} \\
\hline
\multirow{2}{*}{Teens}      & F \hspace{4pt}112 spks & 31.15\% & 31.61\% & 13.82\% & 14.08\% & 15.01\% & 17.66\% \\
                            & M 112 spks & 23.12\% & 19.98\% &  4.53\% &  5.02\% &  6.89\% & 10.22\% \\
\multirow{2}{*}{Twenties}   & F \hspace{4pt}582 spks & 23.27\% & 19.34\% &  3.67\% &  5.41\% &  5.93\% & 11.50\% \\
                            & M 582 spks & 22.95\% & 20.87\% &  6.62\% &  8.16\% &  8.27\% & 11.56\% \\
\multirow{2}{*}{Thirties}   & F \hspace{4pt}240 spks & 22.70\% & 22.34\% &  2.94\% &  4.67\% &  5.52\% & 8.17\% \\
                            & M 240 spks & 20.14\% & 17.05\% &  2.69\% &  3.80\% &  3.99\% & 9.25\% \\
\multirow{2}{*}{Fourties}   & F \hspace{4pt}140 spks & 20.75\% & 19.30\% &  2.47\% &  4.08\% &  4.84\% & 8.47\% \\
                            & M 140 spks & 17.70\% & 17.33\% &  1.79\% &  3.09\% &  3.00\% & 5.82\% \\
\multirow{2}{*}{Fifties}    & F \hspace{4pt}110 spks & 24.49\% & 23.09\% &  2.88\% &  5.54\% &  6.43\% & 18.26\% \\
                            & M 126 spks & 18.99\% & 16.55\% &  1.39\% &  2.68\% &  3.78\% & 12.2\% \\
\multirow{2}{*}{Sixties}    & F \hspace{9pt}49 spks & 27.95\% & 26.11\% &  5.60\% & 11.11\% & 10.63\% & 24.69\% \\
                            & M  \hspace{5pt}57 spks & 20.09\% & 18.27\% &  1.66\% &  4.15\% &  3.71\% & 6.95\% \\
\multirow{2}{*}{Seventy+}   & F  \hspace{9pt}17 spks & 19.61\% & 16.37\% &  1.64\% &  6.56\% &  1.86\% & 5.39\% \\
                            & M  \hspace{5pt}69 spks & 21.09\% & 25.23\% &  8.89\% & 11.96\% &  9.95\% & 12.92\% \\
\hline
\end{tabular}
}
\caption{\label{tab:eer_age_gender}CommonVoice: 
EER variation over age for both genders.
}
\end{table*}
SVeritas evaluates the robustness of SV models through a structured three-stage pipeline: (1) scenario simulation, (2) embedding extraction, and (3) performance evaluation. As illustrated in Figure~\ref{fig:sveritasFig}, the first stage introduces a wide range of real-world and synthetic perturbations to both enrollment and test audio. These include natural variations (e.g., speaking style, duration, and linguistic content), environmental conditions (e.g., noise, reverberation, and microphone distance), and recording artifacts (e.g., codec compression and bandwidth limitations). SVeritas also incorporates physical and demographic variability, such as speaker age, health, and accent, as well as adversarial factors including spoofing attempts and both black-box and white-box attacks. The second stage applies multiple state-of-the-art embedding models to extract speaker representations. Finally, performance is evaluated using metrics such as Equal Error Rate (EER) across matched/mismatched scenarios and demographic subgroups, enabling a comprehensive assessment of model robustness and fairness.

\subsection{Scenario Simulation}
SVeritas evaluates speaker verification systems across a range of real-world and synthetic scenarios. These simulations are organized into six broad categories, each capturing a unique aspect of deployment variability or robustness challenge.

The data themselves were obtained by simulating the various effects on a number of public corpora such as EARS \cite{richter2024ears}, AMI Meeting Corpus \cite{kraaij2005ami} and Mozilla CommonVoice 21 \cite{ardila2020common}. In order to maximally ensure fair implementation of the benchmark we only employed the \textit{test} portions of the corpora, under the assumption that developers of SV systems are unlikely to have used these to train the model.

\subsubsection{Audio Capture}
The \textit{audio capture} benchmark evaluates the performance of SV systems under various audio capture conditions that may be encountered in real life.
\begin{enumerate}[leftmargin=*]
    \item \textit{Broadband clean:} These are 16-bit resolution linear PCM 16khz sampled studio-quality data. In the context of speech processing tasks, this has been the long-standing standard for ``ideal'' recordings.
    \item \textit{G-711:} The G-711 standard data are captured with 8-bit mu law quantization,  sampled at 8khz. These remain common in telephony applications. The G-711 achieves a low bitrate of 64kbps \cite{recommendation1988pulse}.
    % \item \textit{G-729:} The G-729 is a typical low-bitrate codec employed in telephony and VIP applications, operating on telephony bandwith (8khz sampled) signals. The speech is typically compressed through a lossy code-excited linear prediction (CELP) to achieve bitrates as low as 8kbps  and consequently introduces distortions [CITE].
    \item \textit{GSM 06.10:} The GSM 06.10 is a legacy codec, standardized for 2G GSM mobile communications, operates on 8 kHz sampled signals and uses Regular Pulse Excitation with Long Term Prediction (RPE-LTP) to compress speech to approximately 13 kbps, introducing characteristic bandlimited and quantization distortions \cite{tc1993european}.
    \item \textit{Opus:} Opus is a dynamic-bitrate codec employed in applications such as WhatsApp \cite{Kumar2024MLow}, Zoom \cite{zoom_premium_audio} and WebRTC \cite{valin2012definition}  . It dynamically adjusts the compression of the signal according to current network conditions.  SVeritas uses Opus in two modes, narrowband 8khz and wideband 16khz and randomly selects one of the two to apply to any signal, to simulate the unpredictable nature of the compression. 
    \item \textit{AMR:} The ``Adaptive MultiRate'' (AMR) codec is a legacy codec prevalent particularly in 2G and 3G cellular networks. It operates on 8khz mu-law sampled data, and employs variants of CELP coding, but the dynamic switching enables higher-quality audio.  SVeritas chooses randomly between AMR-Narrowband (4.75–12.2 kbps)) and AMR-wideband (12.6 kbps or higher) \cite{sjoberg2007rtp} to emulate the dynamic nature of the codec.
\end{enumerate}
%The CommonVoice 21 corpus was used to generate these data.  
We evaluate systems both under conditions of \textit{match}, where the same codec is used for both test and enrollment data, and \textit{mismatch}, where the two are different. Note that while the results reported in this paper only consider the codecs mentioned above, our actual package implements and tests against a wider set of popular codecs.

\subsubsection{Noise and Channel}
Real-world recordings are often affected by the room responses of the space they are recorded in and any noise sources present in them.  These introduce distortions which may be further exacerbated by coding schemes that are part of the data capture and transmission.  The \textit{noise and channel} benchmark evaluates the robustness of the SV system under these conditions.
\begin{enumerate}[leftmargin=*]
    \item \textit{Noise:} We evaluate the performance under three varities of noise, namely gaussian noise, environmental noise and crosstalk, at three different signal to noise ratios of 5, 15 and 25dB SNR.
    \item \textit{Real Room Response:} We also evaluate the influence of the room response. To implement these, we consider room impulse responses (RIRs) of three different severity levels (in terms of T60 times) drawn randomly from the Room Impulse Response and Noise corpus \cite{ko2017study}. 
\end{enumerate}
The actual benchmarks considers both the room responses and the noises in isolation, and their compounded effect (with RIRs applied on top of the noise). All data are generated through digital simulation of these effects on CommonVoice data. Finally, since codecs too will cause additional distortion of noisy speech,  we also consider the effect of codec compression on signals corrupted by noise and room response.  For the results reported in this we have only considered Opus, G-729 and AMR codecs applied to signals corrupted by medium severity levels of room response and noise; our actual package reports results on the comprehensive set of combinations and their summary statistics.

%======================

%Clean speech serves as the baseline for evaluating speaker verification models under ideal conditions. We use high-quality, studio-grade audio from standard datasets such as EARS \cite{richter2024ears}, AMI Meeting Corpus \cite{kraaij2005ami} and Mozilla CommonVoice \cite{ardila2019common}. These recordings provide reliable reference points for measuring degradation under stress conditions.

\subsubsection{Demographic Variations}
To assess fairness and generalization, we evaluate speaker verification models across demographic groups, including variations in gender, age, ethnicity, and native language. A key focus of this category is cross-lingual robustness: we use CommonVoice \cite{ardila2020common} to test whether models trained primarily on English can correctly verify speakers when they speak in other languages. Since speaker identity is grounded in vocal acoustics, a robust SV model should recognize the same speaker regardless of the spoken language. This evaluation reveals whether models rely too heavily on language-specific cues and whether they generalize across linguistic boundaries. We also include TTS-generated speech conditioned on demographic traits to further probe model behavior under controlled variation. This setup allows us to measure demographic robustness and detect possible bias in model predictions.

\subsubsection{Synthetic and Adversarial}
Real-life deployments are also vulnerable to a variety of attacks. The \textit{synthetic and adversarial} benchmark quantifies this vulnerability.  We consider the following attacks.
\begin{enumerate}[leftmargin=*]
    \item \textit{Synthetic speech:} Here we evaluate the vulnerability of the system to synthetic speech. In all test pairs, both recordings are from the same speaker. In one case both recordings are real, whereas in the other one of the two is synthetic. Ideally the system must accept the former and reject the latter. In this paper we employ CosyVoiceTTS \cite{du2024cosyvoice}, xTTS \cite{coqui_xtts}, and StyleTTS \cite{styletts}  for the synthetic speech; the full benchmark also evaluates other TTS systems.
    \item \textit{Adversarial attack:} We consider adversarial attacks where an imposter attempts to mislead the system.  The test is similar to the synthetic speech attack, except that instead of synthetic speech, we have adversarially modified speech. In this paper we consider two adversarial attacks: a \textit{white-box} (full access to model weights) attack, namely  the Fast Gradient Sign Method (FGSM) \cite{goodfellow2014explaining}, and one \textit{black-box} attack (access restricted to output label only), namely the Fakebob attack \cite{chen2021real}.  The full SVeritas benchmark package also considers other popular attacks.
\end{enumerate}

\subsection{Metrics}
We evaluate the performance of speaker verification systems using three standard metrics: Equal Error Rate (EER), minimum Detection Cost Function (minDCF), and Area Under the Curve (AUC). EER is defined as the point at which the false acceptance rate (FAR) equals the false rejection rate (FRR), providing a balanced indicator of accuracy across operating points. It is used as the primary metric due to its intuitive interpretability. minDCF measures the minimum cost achievable when accounting for application-specific penalties (e.g., a higher cost for FAR in high-security contexts) and thus reflects performance under asymmetric decision costs. AUC, a threshold-independent metric, quantifies the separability between genuine and impostor trials and is particularly sensitive to systemic errors in low-FAR regimes, such as those required in forensic applications. 
\begin{table*}[t]
\centering
\resizebox{\textwidth}{!}{
\begin{tabular}{llcccccc}
\hline
\textbf{Codec} & \textbf{Condition} & \textbf{WavLM-Base} & \textbf{WavLM-Base+} & \textbf{RedimNet} & \textbf{ECAPA} & \textbf{MFA-Conformer} & \textbf{Titanet}\\
\hline
                        & Clean             & 23.05\% & 20.23\% &  4.69\% &  6.13\% &  6.65\% & 4.92\% \\
 \hline
\multirow{4}{*}{GSM}    & No Noise          & 23.09\% & 20.24\% &  4.65\% &  6.15\% &  6.72\% & 4.95\% \\ 
                        & GaussNoise+RIR    & 40.64\% & 36.67\% & 22.46\% & 17.89\% & 27.30\% & 26.01\% \\
                        & EnvNoise+RIR      & 40.32\% & 38.63\% & 17.24\% & 16.47\% & 20.12\% & 22.58\% \\
                        & CrossTalk+RIR     & 40.47\% & 38.29\% & 24.63\% & 24.68\% & 24.95\% & 24.15\% \\
                        \hline
\multirow{4}{*}{AMR}    & No Noise          & 23.09\% & 20.24\% &  4.65\% &  6.15\% &  6.72\% & 4.95\% \\
                        & GaussNoise+RIR    & 40.64\% & 36.67\% & 22.46\% & 17.89\% & 27.30\% & 26.01\% \\
                        & EnvNoise+RIR      & 40.32\% & 38.63\% & 17.24\% & 16.47\% & 20.12\% & 22.58\% \\
                        & CrossTalk+RIR     & 40.47\% & 38.29\% & 24.63\% & 24.68\% & 24.95\% & 24.15\% \\
                        \hline
\multirow{4}{*}{Opus}   & No Noise          & 23.09\% & 20.24\% &  4.65\% &  6.15\% &  6.72\% & 4.95\% \\
                        & GaussNoise+RIR    & 40.64\% & 36.67\% & 22.46\% & 17.89\% & 27.30\% & 26.01\% \\
                        & EnvNoise+RIR      & 40.32\% & 38.63\% & 17.24\% & 16.47\% & 20.12\% & 22.58\% \\
                        & CrossTalk+RIR     & 40.47\% & 38.29\% & 24.63\% & 24.68\% & 24.95\% & 24.15\% \\
                        \hline
\multirow{4}{*}{AMI}   & NearField (F)        & 39.64\% & 	  34.60\% &	12.12\% &	 21.69\% &	 20.63\% & 16.65\% \\
                        & NearField (M)        & 37.92\% &	  38.69\% &	 17.27\% &	 20.27\% &	 22.31\% & 13.05\% \\
                        & FarField (F)        &  47.06\% &	 47.63\% &	 34.96\% &	 36.04\% &	 36.67\% & 33.28\% \\
                        & FarField (M)        &  46.63\% &	 45.39\% &	 34.65\% &	 35.00\% &	 37.65\% & 36.09\% \\
                        \hline

\hline
\end{tabular}
}
\caption{\label{tab:eer_codec} CommonVoice: EERs under audio degradation from codecs and noise conditions. AMI results reflect real-world variability in near-field and far-field social environments.
}
\end{table*}

\section{Evaluation}
We evaluate several state-of-the-art SV models using SVeritas and analyze their robustness across a broad range of challenging scenarios. We further extend this analysis to examine model behavior across various demographic subgroups, including speaker age, language background, ethnicity, and gender. Prior work \cite{hutiri2022bias} has noted the presence of biases in SV systems, and our findings corroborate and expand upon these observations by revealing that disparities in robustness can emerge across subgroups. These results highlight the importance of standardized evaluation under real-world conditions and underscore the utility of SVeritas in advancing fair and reliable speaker verification.

To further quantify fairness and robustness, we conduct a series of pairwise statistical tests across demographic groups using EER as the primary metric. While each group yields a single EER value per model, we leverage the diversity of five models to enable paired comparisons between groups. This design allows us to assess whether performance disparities are consistent across architectures. Full statistical test tables, including $t$-statistics, $p$-values, and significance levels, are provided in Appendix~\ref{sec:appendix-ttests}.
\subsection{Models}
We evaluate a range of SOTA SV models which are publicly available, including WavLM-Base and WavLM-Base+ \cite{chen2022wavlm}, ECAPA-TDNN \cite{desplanques2020ecapa}, Titanet \cite{koluguri2022titanet}, and RedimNet \cite{yakovlev2024reshape}. In addition, we include MFA-Conformer\cite{zhang2022mfa}, which we train ourselves due to the lack of publicly available checkpoints. All publicly available models are sourced from official repositories or HuggingFace implementations, where applicable. We mentioned more information about each model in the Appendix~\ref{sec:appendix-arch}. 

\subsection{Robustness in Noise Environment}
We evaluate robustness to noise and channel variability using two benchmarks: synthetic distortions applied to CommonVoice and real-world social conditions captured in the AMI corpus. For synthetic testing, we simulate Gaussian, environmental, and cross-talk noise at varying SNRs (5, 15, 25 dB) with and without room impulse response (RIR) of severity levels 2, 3, and 4. These are evaluated under three codecs (GSM, AMR, Opus), with results shown in Table 4. For real-world testing, we use AMI recordings captured in near-field and far-field microphone setups to assess model performance in natural interactive environments.

As shown in \ref{tab:eer_codec} and \ref{tab:eer_noise}, we observe a consistent trend: WavLM-based models degrade rapidly in the presence of noise and reverberation, especially when combined with low-bitrate codecs such as GSM or AMR. For instance, WavLM-Base shows a sharp EER increase from 23.05\% in clean conditions to over 40\% across nearly all noisy+RIR combinations. In contrast, RedimNet, ECAPA-TDNN, and MFA-Conformer exhibit significantly stronger robustness, maintaining substantially lower EERs in both synthetic and real-world conditions. These results emphasize the need for channel-aware model development and highlight the importance of including realistic acoustic variation during training.

\subsection{Robustness in Speaking Style (Lombard Condition)}
Lombard speech refers to a natural adaptation in which speakers involuntarily raise vocal intensity, modify articulation, and adjust prosody in the presence of background noise. To assess SV robustness under this condition, we use the Lombard GRID corpus, which includes 54 speakers producing both plain and noise-induced Lombard utterances in constrained GRID syntax. 

We construct exhaustive cosine-similarity trials within gender and condition. For the \textit{Plain} and \textit{Lombard} settings, trials are drawn from within the same condition (e.g., plain–plain, lombard–lombard). For the \textit{Mixed} setting, target trials consist of cross-condition utterances (plain–lombard for the same speaker), while impostor trials remain within-condition. This design quantifies both matched-condition performance and cross-style domain mismatch.

As shown in Appendix Table~\ref{tab:eer_lombard}, domain mismatch is the dominant source of error: all models exhibit elevated EERs under the \textit{Mixed} setting. In contrast, matched Lombard trials are comparable to, and in some cases slightly outperform, plain trials. WavLM-Base and WavLM-Base+ are particularly affected, reaching $\approx$15--20\% EER even under matched conditions, whereas RedimNet, ECAPA-TDNN, MFA-Conformer, and Titanet remain below 2.5\% EER. These findings highlight that self-supervised models are especially vulnerable to cross-style mismatch, while architectures with stronger aggregation mechanisms demonstrate more stable generalization.

\subsection{Robustness in Adversarial Scenarios}
We run different tests for adversarial attacks and TTS spoofing. For the TTS spoofing, we use the EARS dataset, which contains 109 speakers (59 female, 50 male). For Real vs. Synthetic evaluations, both utterances in a pair originate from the same speaker. Positive pairs consist of two real utterances, while negative pairs include one real and one synthetic sample generated by a TTS system. For adversarial attacks, we adopt a targeted verification setup in which the first utterance is adversarially perturbed using either FGSM or FakeBob, and the second is a clean utterance from the same speaker. This design ensures speaker consistency while isolating the effect of the perturbation.

As shown in Table~\ref{tab:merged_conditions}, MFA-Conformer demonstrates the strongest robustness across all tested conditions, achieving 0\% EER under TTS spoofing and the lowest error rates under both FGSM (45.26\%) and FakeBob (35.48\%) attacks. RedimNet also performs well under TTS spoofing (5.94\%), though it is more susceptible to adversarial attacks. In contrast, ECAPA-TDNN is highly vulnerable to FakeBob, reaching an EER of 62.36\%. The WavLM-based models (Base and Base+) show consistent vulnerability under both spoofing and adversarial conditions. These findings highlight substantial variability in model robustness and underscore the need to develop verification systems that are resilient to both synthetic speech and adversarial manipulation.
For completeness, we also include in Appendix table ~\ref{tab:tts_conditions} an expanded table reporting results with additional TTS systems. This ensures that our conclusions are not limited to a small set of spoofing generators, but generalize across a broader range of synthesis pipelines.

\subsection{Robustness in Demographic Variations}
To investigate demographic bias in speaker verification, we generate verification pairs by first splitting the dataset by gender, and then further dividing each gender group based on the desired demographic category such as age, ethnicity, or language. Within each subgroup, we form same-speaker pairs and compute the EER  independently for each model. This stratified evaluation allows us to analyze whether models exhibit bias or performance disparities across demographic dimensions, especially among underrepresented groups. We run these experiments on both the EARS and Mozilla CommonVoice datasets, leveraging their detailed metadata. 

As seen in Table~\ref{tab:demographic_eer}, certain language-gender or ethnicity-gender combinations (e.g., male-Hispanic, female-Asian) have significantly fewer speakers and exhibit elevated EERs, suggesting weaker generalization. Our analysis reveals signs of demographic bias in several models. For instance, WavLM-Base shows degradation in older age groups. In contrast, Redimnet maintains the most consistent performance regarding all demographic splits, with minimal variation across age, gender, and ethnicity. 

Using paired t-tests across five models in the EARS dataset \cite{richter2024ears}, we assess consistency of group-level EER differences. Males show higher EERs than females (17.3\% vs. 13.7\%), but the gap is not statistically significant ($p = 0.095$). Younger males (18–25) outperform older males ($p < 0.05$). Females aged 36–45 significantly outperform other female age groups ($p < 0.01$). Black females show significantly higher EERs than white females ($p < 0.001$). Asian and Hispanic males also perform worse, but sample sizes are small ($n \leq 5$) (see Appendix~\ref{sec:appendix-ears}).

CommonVoice results (Table~\ref{tab:eer_age_gender}, Table~\ref{tab:mindcf_age_gender},  Table~\ref{tab:auc_age_gender}, Appendix~\ref{sec:appendix-commonvoice}) confirm these trends. Gender identity groups show no significant EER gap ($p = 0.779$), but age remains a major factor. Older male and female speakers (60+) consistently underperform compared to younger groups ($p < 0.01$).

These results collectively suggest that demographic imbalance in training data may contribute to uneven generalization and reduced fairness, particularly in age and ethnicity subgroups with limited representation.

\section{Conclusion}
We present \textit{SVeritas}, a comprehensive and extensible benchmark for evaluating speaker verification models under diverse real-world and synthetic stressors. Unlike prior work, it covers environmental noise, channel mismatches, codecs, cross-lingual variation, demographic shifts, adversarial attacks, and importantly, TTS-based spoofing—an often overlooked but growing threat. SVeritas enables fine-grained robustness and fairness analysis across gender, age, ethnicity, and language, while offering a modular framework that supports easy integration of new models and evaluation settings. It provides a unified, reproducible foundation for building SV systems that are not only accurate, but also resilient, equitable, and ready for real-world deployment.

\section*{Limitations}
While \textit{SVeritas} provides a broad and extensible evaluation framework, it currently applies stress conditions at fixed levels of severity. This design simplifies benchmarking and ensures consistency across models, but may not fully capture how systems degrade under progressively harder conditions. In real-world scenarios, distortions such as noise, reverberation, or compression vary in intensity and interact in complex ways. Future work could extend SVeritas with parameterized or continuous stress levels, enabling finer-grained robustness analysis and stress-adaptive training strategies

\section*{Acknowledgments}
This work was partially funded by Portuguese national funds through Fundação para a Ciência e a Tecnologia (FCT), under project UIDB/50021/2020 (DOI:10.54499/UIDB/50021/2020), and by the Portuguese Recovery and Resilience Plan and NextGenerationEU European Union funds under project C644865762-00000008 (Accelerat.AI).
\bibliography{main}

\begin{thebibliography}{56}
\providecommand{\natexlab}[1]{#1}

\bibitem[{Abdullah et~al.(2025)Abdullah, Badawi, Abdullah, Hamad, Taher, Muhamad, Ahmed, Hassan, Aula, and Rashid}]{abdullah2025dialect}
Abdulhady~Abas Abdullah, Soran Badawi, Dana~A Abdullah, Dana~Rasul Hamad, Hanan~Abdulrahman Taher, Sabat~Salih Muhamad, Aram~Mahmood Ahmed, Bryar~A Hassan, Sirwan~Abdolwahed Aula, and Tarik~A Rashid. 2025.
\newblock From dialect gaps to identity maps: Tackling variability in speaker verification.
\newblock \emph{arXiv preprint arXiv:2505.04629}.

\bibitem[{AI(2023)}]{coqui_xtts}
Coqui AI. 2023.
\newblock Coqui tts: xtts - cross-lingual neural text-to-speech.
\newblock \url{https://github.com/coqui-ai/TTS}.

\bibitem[{Al-Karawi(2021)}]{al2021mitigate}
Khamis~A Al-Karawi. 2021.
\newblock Mitigate the reverberation effect on the speaker verification performance using different methods.
\newblock \emph{International Journal of Speech Technology}, 24(1):143--153.

\bibitem[{Alzantot et~al.(2018)Alzantot, Balaji, and Srivastava}]{alzantot2018did}
Moustafa Alzantot, Bharathan Balaji, and Mani Srivastava. 2018.
\newblock Did you hear that? adversarial examples against automatic speech recognition.
\newblock \emph{arXiv preprint arXiv:1801.00554}.

\bibitem[{Ardila et~al.(2020)Ardila, Branson, Davis, Kohler, Meyer, Henretty, Morais, Saunders, Tyers, and Weber}]{ardila2020common}
Rosana Ardila, Megan Branson, Kelly Davis, Michael Kohler, Josh Meyer, Michael Henretty, Reuben Morais, Lindsay Saunders, Francis Tyers, and Gregor Weber. 2020.
\newblock Common voice: A massively-multilingual speech corpus.
\newblock In \emph{Proceedings of the Twelfth Language Resources and Evaluation Conference}, pages 4218--4222.

\bibitem[{CCITT(1988)}]{recommendation1988pulse}
CCITT. 1988.
\newblock Pulse code modulation (pcm) of voice frequencies.
\newblock In \emph{ITU}.

\bibitem[{Chen et~al.(2021)Chen, Chenb, Fan, Du, Zhao, Song, and Liu}]{chen2021real}
Guangke Chen, Sen Chenb, Lingling Fan, Xiaoning Du, Zhe Zhao, Fu~Song, and Yang Liu. 2021.
\newblock Who is real bob? adversarial attacks on speaker recognition systems.
\newblock In \emph{2021 IEEE Symposium on Security and Privacy (SP)}, pages 694--711. IEEE.

\bibitem[{Chen et~al.(2022)Chen, Wang, Chen, Wu, Liu, Chen, Li, Kanda, Yoshioka, Xiao et~al.}]{chen2022wavlm}
Sanyuan Chen, Chengyi Wang, Zhengyang Chen, Yu~Wu, Shujie Liu, Zhuo Chen, Jinyu Li, Naoyuki Kanda, Takuya Yoshioka, Xiong Xiao, and 1 others. 2022.
\newblock Wavlm: Large-scale self-supervised pre-training for full stack speech processing.
\newblock \emph{IEEE Journal of Selected Topics in Signal Processing}, 16(6):1505--1518.

\bibitem[{Dehak et~al.(2010)Dehak, Kenny, Dehak, Dumouchel, and Ouellet}]{dehak2010front}
Najim Dehak, Patrick~J Kenny, R{\'e}da Dehak, Pierre Dumouchel, and Pierre Ouellet. 2010.
\newblock Front-end factor analysis for speaker verification.
\newblock \emph{IEEE Transactions on Audio, Speech, and Language Processing}, 19(4):788--798.

\bibitem[{Desplanques et~al.(2020)Desplanques, Thienpondt, and Demuynck}]{desplanques2020ecapa}
Brecht Desplanques, Jenthe Thienpondt, and Kris Demuynck. 2020.
\newblock Ecapa-tdnn: Emphasized channel attention, propagation and aggregation in tdnn based speaker verification.
\newblock \emph{Interspeech}.

\bibitem[{Dey(2018)}]{dey2018phonetic}
Subhadeep Dey. 2018.
\newblock Phonetic aware techniques for speaker verification.
\newblock Technical report, EPFL.

\bibitem[{Doddington(2012)}]{doddington2012effect}
George~R Doddington. 2012.
\newblock The effect of target/non-target age difference on speaker recognition performance.
\newblock In \emph{Odyssey}, pages 263--267.

\bibitem[{Du et~al.(2024)Du, Chen, Zhang, Hu, Lu, Yang, Hu, Zheng, Gu, Ma et~al.}]{du2024cosyvoice}
Zhihao Du, Qian Chen, Shiliang Zhang, Kai Hu, Heng Lu, Yexin Yang, Hangrui Hu, Siqi Zheng, Yue Gu, Ziyang Ma, and 1 others. 2024.
\newblock Cosyvoice: A scalable multilingual zero-shot text-to-speech synthesizer based on supervised semantic tokens.
\newblock \emph{arXiv preprint arXiv:2407.05407}.

\bibitem[{Goodfellow et~al.(2014)Goodfellow, Shlens, and Szegedy}]{goodfellow2014explaining}
Ian~J Goodfellow, Jonathon Shlens, and Christian Szegedy. 2014.
\newblock Explaining and harnessing adversarial examples.
\newblock \emph{arXiv preprint arXiv:1412.6572}.

\bibitem[{Hajavi and Etemad(2023)}]{hajavi2023study}
Amirhossein Hajavi and Ali Etemad. 2023.
\newblock A study on bias and fairness in deep speaker recognition.
\newblock In \emph{ICASSP 2023-2023 IEEE International Conference on Acoustics, Speech and Signal Processing (ICASSP)}, pages 1--5. IEEE.

\bibitem[{Hautam{\"a}ki et~al.(2013)Hautam{\"a}ki, Kinnunen, Hautam{\"a}ki, Leino, and Laukkanen}]{hautamaki2013vectors}
Rosa~Gonz{\'a}lez Hautam{\"a}ki, Tomi Kinnunen, Ville Hautam{\"a}ki, Timo Leino, and Anne-Maria Laukkanen. 2013.
\newblock I-vectors meet imitators: on vulnerability of speaker verification systems against voice mimicry.
\newblock In \emph{Interspeech}, pages 930--934. Citeseer.

\bibitem[{Hintz and Siegert(2024)}]{hintzcommonbench}
Jan Hintz and Ingo Siegert. 2024.
\newblock Commonbench: A larger scale speaker verification benchmark.
\newblock In \emph{Proc. SPSC 2024}, pages 17--20.

\bibitem[{Hutiri and Ding(2022)}]{hutiri2022bias}
Wiebke~Toussaint Hutiri and Aaron~Yi Ding. 2022.
\newblock Bias in automated speaker recognition.
\newblock In \emph{Proceedings of the 2022 ACM conference on fairness, accountability, and transparency}, pages 230--247.

\bibitem[{Inoue and Goto(2020)}]{inoue2020semi}
Nakamasa Inoue and Keita Goto. 2020.
\newblock Semi-supervised contrastive learning with generalized contrastive loss and its application to speaker recognition.
\newblock In \emph{2020 Asia-Pacific Signal and Information Processing Association Annual Summit and Conference (APSIPA ASC)}, pages 1641--1646. IEEE.

\bibitem[{Jati et~al.(2021)Jati, Hsu, Pal, Peri, AbdAlmageed, and Narayanan}]{jati2021adversarial}
Arindam Jati, Chin-Cheng Hsu, Monisankha Pal, Raghuveer Peri, Wael AbdAlmageed, and Shrikanth Narayanan. 2021.
\newblock Adversarial attack and defense strategies for deep speaker recognition systems.
\newblock \emph{Computer Speech \& Language}, 68:101199.

\bibitem[{Javed et~al.(2023)Javed, Bhogale, Raman, Kumar, Kunchukuttan, and Khapra}]{javed2023indicsuperb}
Tahir Javed, Kaushal Bhogale, Abhigyan Raman, Pratyush Kumar, Anoop Kunchukuttan, and Mitesh~M Khapra. 2023.
\newblock Indicsuperb: a speech processing universal performance benchmark for indian languages.
\newblock In \emph{Proceedings of the Thirty-Seventh AAAI Conference on Artificial Intelligence and Thirty-Fifth Conference on Innovative Applications of Artificial Intelligence and Thirteenth Symposium on Educational Advances in Artificial Intelligence}, pages 12942--12950.

\bibitem[{Kelly and Harte(2011)}]{kelly2011effects}
Finnian Kelly and Naomi Harte. 2011.
\newblock Effects of long-term ageing on speaker verification.
\newblock In \emph{European Workshop on Biometrics and Identity Management}, pages 113--124. Springer.

\bibitem[{Kenny(2010)}]{kenny2010bayesian}
Patrick Kenny. 2010.
\newblock Bayesian speaker verification with heavy-tailed priors.
\newblock In \emph{Proc. Odyssey 2010}, pages paper--14.

\bibitem[{Ko et~al.(2017)Ko, Peddinti, Povey, Seltzer, and Khudanpur}]{ko2017study}
Tom Ko, Vijayaditya Peddinti, Daniel Povey, Michael~L Seltzer, and Sanjeev Khudanpur. 2017.
\newblock A study on data augmentation of reverberant speech for robust speech recognition.
\newblock In \emph{2017 IEEE international conference on acoustics, speech and signal processing (ICASSP)}, pages 5220--5224. IEEE.

\bibitem[{Koluguri et~al.(2022)Koluguri, Park, and Ginsburg}]{koluguri2022titanet}
Nithin~Rao Koluguri, Taejin Park, and Boris Ginsburg. 2022.
\newblock Titanet: Neural model for speaker representation with 1d depth-wise separable convolutions and global context.
\newblock In \emph{ICASSP 2022-2022 IEEE international conference on acoustics, speech and signal processing (ICASSP)}, pages 8102--8106. IEEE.

\bibitem[{Kraaij et~al.(2005)Kraaij, Hain, Lincoln, and Post}]{kraaij2005ami}
Wessel Kraaij, Thomas Hain, Mike Lincoln, and Wilfried Post. 2005.
\newblock The ami meeting corpus.
\newblock In \emph{Proc. International Conference on Methods and Techniques in Behavioral Research}, pages 1--4.

\bibitem[{Kumar et~al.(2024)Kumar, Agarwalla, Srinivasan, Hor, and Wong}]{Kumar2024MLow}
Jatin Kumar, Bikash Agarwalla, Sriram Srinivasan, King~Wei Hor, and Tim Wong. 2024.
\newblock \href {https://engineering.fb.com/2024/06/13/web/mlow-metas-low-bitrate-audio-codec/} {{MLow: Meta’s low bitrate audio codec}}.
\newblock Engineering at Meta.
\newblock Accessed: 2025-05-20.

\bibitem[{Lin et~al.(2024)Lin, Qin, Zhao, Cheng, Jiang, Wu, and Li}]{lin2024voxblink}
Yuke Lin, Xiaoyi Qin, Guoqing Zhao, Ming Cheng, Ning Jiang, Haiying Wu, and Ming Li. 2024.
\newblock Voxblink: A large scale speaker verification dataset on camera.
\newblock In \emph{ICASSP}, pages 10271--10275. IEEE.

\bibitem[{Liu et~al.(2022)Liu, Wang, Ren, Chen, Liu, and Zhao}]{styletts}
Jinglin Liu, Chengyi Wang, Yi~Ren, Feiyang Chen, Peng Liu, and Zhou Zhao. 2022.
\newblock Styletts: A style-based generative model for natural and diverse text-to-speech synthesis.
\newblock \emph{Advances in Neural Information Processing Systems (NeurIPS)}.

\bibitem[{Mo{\v{s}}ner et~al.(2022)Mo{\v{s}}ner, Plchot, Burget, and {\v{C}}ernock{\`y}}]{movsner2022multisv}
Ladislav Mo{\v{s}}ner, Old{\v{r}}ich Plchot, Luk{\'a}{\v{s}} Burget, and Jan~Honza {\v{C}}ernock{\`y}. 2022.
\newblock Multisv: Dataset for far-field multi-channel speaker verification.
\newblock In \emph{ICASSP 2022-2022 IEEE international conference on acoustics, speech and signal processing (ICASSP)}, pages 7977--7981. IEEE.

\bibitem[{Nagrani et~al.(2020)Nagrani, Chung, Huh, Brown, Coto, Xie, McLaren, Reynolds, and Zisserman}]{nagrani2020voxsrc}
Arsha Nagrani, Joon~Son Chung, Jaesung Huh, Andrew Brown, Ernesto Coto, Weidi Xie, Mitchell McLaren, Douglas~A Reynolds, and Andrew Zisserman. 2020.
\newblock Voxsrc 2020: The second voxceleb speaker recognition challenge.
\newblock \emph{arXiv preprint arXiv:2012.06867}.

\bibitem[{Nakamura et~al.(2008)Nakamura, Iwano, and Furui}]{nakamura2008differences}
Masanobu Nakamura, Koji Iwano, and Sadaoki Furui. 2008.
\newblock Differences between acoustic characteristics of spontaneous and read speech and their effects on speech recognition performance.
\newblock \emph{Computer Speech \& Language}, 22(2):171--184.

\bibitem[{Nam et~al.(2024)Nam, Heo, Jung, and Chung}]{nam2024disentangled}
KiHyun Nam, Hee-Soo Heo, Jee-weon Jung, and Joon~Son Chung. 2024.
\newblock Disentangled representation learning for environment-agnostic speaker recognition.
\newblock \emph{arXiv preprint arXiv:2406.14559}.

\bibitem[{Nandwana et~al.(2018)Nandwana, van Hout, McLaren, Stauffer, Richey, Lawson, and Graciarena}]{nandwana2018robust}
Mahesh~Kumar Nandwana, Julien van Hout, Mitchell McLaren, Allen~R Stauffer, Colleen Richey, Aaron Lawson, and Martin Graciarena. 2018.
\newblock Robust speaker recognition from distant speech under real reverberant environments using speaker embeddings.
\newblock In \emph{Interspeech}, pages 1106--1110.

\bibitem[{Njegovec(2025)}]{njegovec2025forensic}
Vendela Njegovec. 2025.
\newblock Forensic automatic speaker recognition: Analyzing codecs for calibration and their impact on system performance.
\newblock Master's thesis, University of Twente.

\bibitem[{Poddar et~al.(2018)Poddar, Sahidullah, and Saha}]{poddar2018speaker}
Arnab Poddar, Md~Sahidullah, and Goutam Saha. 2018.
\newblock Speaker verification with short utterances: a review of challenges, trends and opportunities.
\newblock \emph{IET Biometrics}, 7(2):91--101.

\bibitem[{Polacky et~al.(2016)Polacky, Pocta, and Jarina}]{polacky2016impact}
Jozef Polacky, Peter Pocta, and Roman Jarina. 2016.
\newblock An impact of wideband speech codec mismatch on a performance of gmm-ubm speaker verification over telecommunication channel.
\newblock In \emph{2016 ELEKTRO}, pages 77--82. IEEE.

\bibitem[{Qin et~al.(2020)Qin, Li, Bu, Rao, Das, Narayanan, and Li}]{qin2020interspeech}
Xiaoyi Qin, Ming Li, Hui Bu, Wei Rao, Rohan~Kumar Das, Shrikanth Narayanan, and Haizhou Li. 2020.
\newblock The interspeech 2020 far-field speaker verification challenge.
\newblock \emph{arXiv preprint arXiv:2005.08046}.

\bibitem[{Reynolds et~al.(2000)Reynolds, Quatieri, and Dunn}]{reynolds2000speaker}
Douglas~A Reynolds, Thomas~F Quatieri, and Robert~B Dunn. 2000.
\newblock Speaker verification using adapted gaussian mixture models.
\newblock \emph{Digital signal processing}, 10(1-3):19--41.

\bibitem[{Richter et~al.(2024)Richter, Wu, Krenn, Welker, Lay, Watanabe, Richard, and Gerkmann}]{richter2024ears}
Julius Richter, Yi-Chiao Wu, Steven Krenn, Simon Welker, Bunlong Lay, Shinjii Watanabe, Alexander Richard, and Timo Gerkmann. 2024.
\newblock {EARS}: An anechoic fullband speech dataset benchmarked for speech enhancement and dereverberation.
\newblock In \emph{Interspeech}.

\bibitem[{Sadjadi et~al.(2022)Sadjadi, Greenberg, Singer, Mason, and Reynolds}]{sadjadi20222021}
Seyed~Omid Sadjadi, Craig Greenberg, Elliot Singer, Lisa Mason, and Douglas Reynolds. 2022.
\newblock The 2021 nist speaker recognition evaluation.
\newblock \emph{arXiv preprint arXiv:2204.10242}.

\bibitem[{Sadjadi et~al.(2017)Sadjadi, Kheyrkhah, Tong, Greenberg, Reynolds, Singer, Mason, and Hernandez-Cordero}]{sadjadi20172016}
Seyed~Omid Sadjadi, Timoth{\'e}e Kheyrkhah, Audrey Tong, Craig Greenberg, Douglas Reynolds, Elliot Singer, Lisa Mason, and Jaime Hernandez-Cordero. 2017.
\newblock The 2016 nist speaker recognition evaluation.
\newblock \emph{Interspeech 2017}.

\bibitem[{Shah et~al.(2025)Shah, Noguero, Heikkila, Raj, and Kourtellis}]{shah2024speech}
Muhammad~A Shah, David~Solans Noguero, Mikko~A Heikkila, Bhiksha Raj, and Nicolas Kourtellis. 2025.
\newblock Speech robust bench: a robustness benchmark for speech recognition.
\newblock \emph{ICLR}.

\bibitem[{Sjoberg et~al.(2007)Sjoberg, Westerlund, Lakaniemi, and Xie}]{sjoberg2007rtp}
Johan Sjoberg, Magnus Westerlund, Ari Lakaniemi, and Qiaobing Xie. 2007.
\newblock Rtp payload format and file storage format for the adaptive multi-rate (amr) and adaptive multi-rate wideband (amr-wb) audio codecs.
\newblock Technical report, IETF.

\bibitem[{TC-SMG(1993)}]{tc1993european}
ETSI TC-SMG. 1993.
\newblock European digital cellular telecommunications system (phase 2); full rate speech transcoding (gsm 06.10).
\newblock \emph{European Telecommunication Standard ETS}, 300:580--2.

\bibitem[{Thienpondt and Demuynck(2023)}]{thienpondt2023ecapa2}
Jenthe Thienpondt and Kris Demuynck. 2023.
\newblock Ecapa2: A hybrid neural network architecture and training strategy for robust speaker embeddings.
\newblock In \emph{2023 IEEE Automatic Speech Recognition and Understanding Workshop (ASRU)}, pages 1--8. IEEE.

\bibitem[{Valin et~al.(2012)Valin, Vos, and Terriberry}]{valin2012definition}
Jean-Marc Valin, Koen Vos, and Timothy Terriberry. 2012.
\newblock Definition of the opus audio codec.
\newblock Technical report, IETF.

\bibitem[{Villalba et~al.(2020)Villalba, Chen, Snyder, Garcia-Romero, McCree, Sell, Borgstrom, Garc{\'\i}a-Perera, Richardson, Dehak et~al.}]{villalba2020state}
Jes{\'u}s Villalba, Nanxin Chen, David Snyder, Daniel Garcia-Romero, Alan McCree, Gregory Sell, Jonas Borgstrom, Leibny~Paola Garc{\'\i}a-Perera, Fred Richardson, R{\'e}da Dehak, and 1 others. 2020.
\newblock State-of-the-art speaker recognition with neural network embeddings in nist sre18 and speakers in the wild evaluations.
\newblock \emph{Computer Speech \& Language}, 60:101026.

\bibitem[{Wu et~al.(2017)Wu, Yamagishi, Kinnunen, Hanil{\c{c}}i, Sahidullah, Sizov, Evans, Todisco, and Delgado}]{wu2017asvspoof}
Zhizheng Wu, Junichi Yamagishi, Tomi Kinnunen, Cemal Hanil{\c{c}}i, Mohammed Sahidullah, Aleksandr Sizov, Nicholas Evans, Massimiliano Todisco, and Hector Delgado. 2017.
\newblock Asvspoof: the automatic speaker verification spoofing and countermeasures challenge.
\newblock \emph{IEEE Journal of Selected Topics in Signal Processing}, 11(4):588--604.

\bibitem[{Yakovlev et~al.(2024)Yakovlev, Makarov, Balykin, Malov, Okhotnikov, and Torgashov}]{yakovlev2024reshape}
Ivan Yakovlev, Rostislav Makarov, Andrei Balykin, Pavel Malov, Anton Okhotnikov, and Nikita Torgashov. 2024.
\newblock Reshape dimensions network for speaker recognition.
\newblock In \emph{Proc. Interspeech 2024}, pages 3235--3239.

\bibitem[{Yang et~al.(2021)Yang, Chi, Chuang, Lai, Lakhotia, Lin, Liu, Shi, Chang, Lin et~al.}]{yang2021superb}
Shu-wen Yang, Po-Han Chi, Yung-Sung Chuang, Cheng-I~Jeff Lai, Kushal Lakhotia, Yist~Y Lin, Andy~T Liu, Jiatong Shi, Xuankai Chang, Guan-Ting Lin, and 1 others. 2021.
\newblock Superb: Speech processing universal performance benchmark.
\newblock \emph{Interspeech}.

\bibitem[{Zeinali et~al.(2019)Zeinali, Lee, Alam, and Burget}]{zeinali2019short}
Hossein Zeinali, Kong~Aik Lee, Jahangir Alam, and Lukas Burget. 2019.
\newblock Short-duration speaker verification (sdsv) challenge 2021: the challenge evaluation plan.
\newblock \emph{arXiv preprint arXiv:1912.06311}.

\bibitem[{Zhang et~al.(2022)Zhang, Lv, Wu, Zhang, Hu, Wu, Lee, and Meng}]{zhang2022mfa}
Yang Zhang, Zhiqiang Lv, Haibin Wu, Shanshan Zhang, Pengfei Hu, Zhiyong Wu, Hung-yi Lee, and Helen Meng. 2022.
\newblock Mfa-conformer: Multi-scale feature aggregation conformer for automatic speaker verification.
\newblock In \emph{Proc. Interspeech 2022}, pages 306--310.

\bibitem[{Zhou et~al.(2023)Zhou, Chen, Wang, Li, and Wang}]{zhou2023adversarial}
Zhenyu Zhou, Junhui Chen, Namin Wang, Lantian Li, and Dong Wang. 2023.
\newblock Adversarial data augmentation for robust speaker verification.
\newblock In \emph{Proceedings of the 2023 9th International Conference on Communication and Information Processing}, pages 226--230.

\bibitem[{{Zoom Video Communications, Inc.}(n.d.)}]{zoom_premium_audio}
{Zoom Video Communications, Inc.} n.d.
\newblock \href {https://explore.zoom.us/docs/doc/Zoom-Premium-Audio.pdf} {\emph{Premium Audio}}.
\newblock Zoom.
\newblock PDF.

\bibitem[{Zuo et~al.(2024)Zuo, Jia, and Li}]{zuo2024advtts}
Chu-Xiao Zuo, Zhi-Jun Jia, and Wu-Jun Li. 2024.
\newblock Advtts: Adversarial text-to-speech synthesis attack on speaker identification systems.
\newblock In \emph{ICASSP 2024-2024 IEEE International Conference on Acoustics, Speech and Signal Processing (ICASSP)}, pages 4840--4844. IEEE.

\end{thebibliography}

\appendix
\onecolumn  % Force single-column layout
\section{Pairwise Statistical Tests}
\label{sec:appendix-ttests}

Each demographic group has a single EER value per model. While speaker counts are sometimes small (as few as 2–3), the number of utterances per speaker is high, resulting in relatively stable EER estimates per group. Traditional t-tests require multiple observations per group; instead, we leverage the fact that we have five models and treat their EERs as paired samples. 

We apply a paired t-test to assess whether models consistently yield higher EERs for one group than another. This test evaluates the consistency of performance differences across models, not population-level differences. We report the number of speakers per group and disregard comparisons involving statistically underpowered cases.

Each table below compares pairwise group performance:
\begin{itemize}
  \item \textbf{Rows:} Reference groups (with speaker count).
  \item \textbf{Columns:} Comparison groups (with speaker count).
  \item \textbf{Cells:} \texttt{t-stat / p-value / significance}
        \begin{itemize}
          \item $t$-stat $< 0$: comparison group has \textit{higher} EER than reference.
          \item $t$-stat $> 0$: comparison group has \textit{lower} EER than reference.
          \item Significance levels: \textbf{***} for $p < 0.001$ (very highly significant); \textbf{**} for $p < 0.01$ (highly significant); \textbf{*} for $p < 0.05$ (statistically significant); No star for $p \geq 0.05$ (not significant).
        \end{itemize}
\end{itemize}

\subsection{EARS Dataset \cite{richter2024ears}}
\label{sec:appendix-ears}

\begin{table}[H]
\centering
\begin{tabular}{lcc}
\hline
 & F (n=59) & M (n=43) \\
\hline
F (n=59) & --- & -2.18 / 0.095 / \\
M (n=43) & 2.18 / 0.095 / & --- \\
\hline
\end{tabular}
\caption{\label{tab:gender}EARS: Pairwise t-tests for gender groups (t / p / sig).}
\end{table}

\begin{table}[H]
\centering
\resizebox{\textwidth}{!}{
\begin{tabular}{lccccc}
\hline
 & M\_18-25 (n=14) & M\_26-35 (n=10) & M\_36-45 (n=10) & M\_46-55 (n=4) & M\_56-65 (n=5) \\
\hline
M\_18-25 (n=14) & --- & 3.70 / 0.021 / * & 3.87 / 0.018 / * & 0.34 / 0.751 / & -0.45 / 0.678 / \\
M\_26-35 (n=10) & -3.70 / 0.021 / * & --- & 2.43 / 0.072 / & -2.89 / 0.044 / * & -2.32 / 0.081 / \\
M\_36-45 (n=10) & -3.87 / 0.018 / * & -2.43 / 0.072 / & --- & -3.04 / 0.038 / * & -2.47 / 0.069 / \\
M\_46-55 (n=4)  & -0.34 / 0.751 / & 2.89 / 0.044 / * & 3.04 / 0.038 / * & --- & -0.78 / 0.478 / \\
M\_56-65 (n=5)  & 0.45 / 0.678 / & 2.32 / 0.081 / & 2.47 / 0.069 / & 0.78 / 0.478 / & --- \\
\hline
\end{tabular}
}
\caption{\label{tab:male-age}EARS: Pairwise t-tests for male age groups (t / p / sig).}
\end{table}

\begin{table}[h]
\centering
\resizebox{\textwidth}{!}{
\begin{tabular}{lcccccc}
\hline
 & F\_18-25 (n=13) & F\_26-35 (n=13) & F\_36-45 (n=7) & F\_46-55 (n=14) & F\_56-65 (n=10) & F\_66-75 (n=2) \\
\hline
F\_18-25 (n=13) & --- & 0.17 / 0.871 / & 4.88 / 0.008 / ** & 0.94 / 0.398 / & -0.04 / 0.967 / & -0.78 / 0.481 / \\
F\_26-35 (n=13) & -0.17 / 0.871 / & --- & 5.15 / 0.007 / ** & 6.02 / 0.004 / ** & -0.32 / 0.762 / & -1.12 / 0.327 / \\
F\_36-45 (n=7)  & -4.88 / 0.008 / ** & -5.15 / 0.007 / ** & --- & -4.54 / 0.010 / * & -2.58 / 0.061 / & -1.93 / 0.126 / \\
F\_46-55 (n=14) & -0.94 / 0.398 / & -6.02 / 0.004 / ** & 4.54 / 0.010 / * & --- & -1.24 / 0.283 / & -1.37 / 0.243 / \\
F\_56-65 (n=10) & 0.04 / 0.967 / & 0.32 / 0.762 / & 2.58 / 0.061 / & 1.24 / 0.283 / & --- & -1.34 / 0.252 / \\
F\_66-75 (n=2)  & 0.78 / 0.481 / & 1.12 / 0.327 / & 1.93 / 0.126 / & 1.37 / 0.243 / & 1.34 / 0.252 / & --- \\
\hline
\end{tabular}
}
\caption{\label{tab:female-age}EARS: Pairwise t-tests for female age groups (t / p / sig).}
\end{table}

\begin{table}[H]
\centering
\resizebox{\textwidth}{!}{
\begin{tabular}{lcccccccc}
\hline
 & F\_white (n=40) & F\_black (n=13) & F\_asian (n=2) & F\_hispanic (n=4) & M\_white (n=31) & M\_black (n=5) & M\_asian (n=2) & M\_hispanic (n=5) \\
\hline
F\_white (n=40) & --- & -9.76 / 0.001 / *** & 1.32 / 0.257 / & 2.73 / 0.053 / & -2.29 / 0.084 / & -0.78 / 0.477 / & 1.27 / 0.274 / & -2.30 / 0.083 / \\
F\_black (n=13) & 9.76 / 0.001 / *** & --- & 1.79 / 0.148 / & 3.57 / 0.023 / * & -0.69 / 0.527 / & 0.80 / 0.466 / & 2.19 / 0.094 / & -1.22 / 0.289 / \\
F\_asian (n=2)  & -1.32 / 0.257 / & -1.79 / 0.148 / & --- & 0.10 / 0.923 / & -1.62 / 0.180 / & -1.20 / 0.295 / & -0.56 / 0.608 / & -1.72 / 0.161 / \\
F\_hispanic (n=4) & -2.73 / 0.053 / & -3.57 / 0.023 / * & -0.10 / 0.923 / & --- & -2.62 / 0.059 / & -2.02 / 0.113 / & -0.91 / 0.415 / & -2.56 / 0.062 / \\
M\_white (n=31) & 2.29 / 0.084 / & 0.69 / 0.527 / & 1.62 / 0.180 / & 2.62 / 0.059 / & --- & 1.73 / 0.159 / & 4.52 / 0.011 / * & -1.88 / 0.133 / \\
M\_black (n=5)  & 0.78 / 0.477 / & -0.80 / 0.466 / & 1.20 / 0.295 / & 2.02 / 0.113 / & -1.73 / 0.159 / & --- & 1.72 / 0.161 / & -2.32 / 0.081 / \\
M\_asian (n=2)  & -1.27 / 0.274 / & -2.19 / 0.094 / & 0.56 / 0.608 / & 0.91 / 0.415 / & -4.52 / 0.011 / * & -1.72 / 0.161 / & --- & -4.09 / 0.015 / * \\
M\_hispanic (n=5) & 2.30 / 0.083 / & 1.22 / 0.289 / & 1.72 / 0.161 / & 2.56 / 0.062 / & 1.88 / 0.133 / & 2.32 / 0.081 / & 4.09 / 0.015 / * & --- \\
\hline
\end{tabular}
}
\caption{\label{tab:ethnicity}EARS: Pairwise t-tests for ethnicity groups (t / p / sig).}
\end{table}

\FloatBarrier
\subsection{CommonVoice Dataset \cite{ardila2020common}}
\label{sec:appendix-commonvoice}

\begin{table}[H]
\centering
\begin{tabular}{lcc}
\hline
 & female\_feminine (n=45) & male\_masculine (n=21) \\
\hline
female\_feminine (n=45) & --- & 0.30 / 0.779 / \\
male\_masculine (n=21)  & -0.30 / 0.779 / & --- \\
\hline
\end{tabular}
\caption{\label{tab:cv-gender}CommonVoice: Pairwise t-tests for gender identity groups (t / p / sig).}
\end{table}

\begin{table}[h]
\centering
\resizebox{\textwidth}{!}{
\begin{tabular}{lccccccc}
\hline
 & M\_teens (n=112) & M\_twenties (n=582) & M\_thirties (n=240) & M\_fourties (n=140) & M\_fifties (n=126) & M\_sixties (n=57) & M\_seventy\_plus (n=69) \\
\hline
M\_teens (n=112) & --- & -2.62/0.059/ & 6.61/0.003/** & 5.44/0.006/** & 11.14/0.000/*** & 5.21/0.006/** & -2.31/0.082/ \\
M\_twenties (n=582) & 2.62/0.059/ & --- & 13.86/0.000/*** & 14.78/0.000/*** & 16.52/0.000/*** & 8.21/0.001/** & -1.88/0.133/ \\
M\_thirties (n=240) & -6.61/0.003/** & -13.86/0.000/*** & --- & 2.18/0.095/ & 4.03/0.016/* & -0.12/0.909/ & -4.47/0.011/* \\
M\_fourties (n=140) & -5.44/0.006/** & -14.78/0.000/*** & -2.18/0.095/ & --- & -0.24/0.821/ & -2.45/0.070/ & -7.39/0.002/** \\
M\_fifties (n=126) & -11.14/0.000/*** & -16.52/0.000/*** & -4.03/0.016/* & 0.24/0.821/ & --- & -2.63/0.058/ & -5.29/0.006/** \\
M\_sixties (n=57) & -5.21/0.006/** & -8.21/0.001/** & 0.12/0.909/ & 2.45/0.070/ & 2.63/0.058/ & --- & -4.73/0.009/** \\
M\_seventy\_plus (n=69) & 2.31/0.082/ & 1.88/0.133/ & 4.47/0.011/* & 7.39/0.002/** & 5.29/0.006/** & 4.73/0.009/** & --- \\
\hline
\end{tabular}
}
\caption{\label{tab:cv-male-age}CommonVoice: Pairwise t-tests for male age groups (t / p / sig).}
\end{table}

\begin{table}[H]
\centering
\resizebox{\textwidth}{!}{
\begin{tabular}{lccccccc}
\hline
 & F\_teens (n=112) & F\_twenties (n=582) & F\_thirties (n=240) & F\_fourties (n=140) & F\_fifties (n=110) & F\_sixties (n=49) & F\_seventy\_plus (n=17) \\
\hline
F\_teens (n=112) & --- & 12.67/0.000/*** & 24.19/0.000/*** & 25.00/0.000/*** & 12.72/0.000/*** & 5.07/0.007/** & 9.42/0.001/*** \\
F\_twenties (n=582) & -12.67/0.000/*** & --- & -0.15/0.885/ & 3.13/0.035/* & -1.25/0.279/ & -5.92/0.004/** & 2.48/0.068/ \\
F\_thirties (n=240) & -24.19/0.000/*** & 0.15/0.885/ & --- & 2.69/0.055/ & -2.89/0.045/* & -7.12/0.002/** & 1.85/0.138/ \\
F\_fourties (n=140) & -25.00/0.000/*** & -3.13/0.035/* & -2.69/0.055/ & --- & -3.27/0.031/* & -7.92/0.001/** & 1.09/0.338/ \\
F\_fifties (n=110) & -12.72/0.000/*** & 1.25/0.279/ & 2.89/0.045/* & 3.27/0.031/* & --- & -7.45/0.002/** & 2.36/0.078/ \\
F\_sixties (n=49) & -5.07/0.007/** & 5.92/0.004/** & 7.12/0.002/** & 7.92/0.001/** & 7.45/0.002/** & --- & 6.02/0.004/** \\
F\_seventy\_plus (n=17) & -9.42/0.001/*** & -2.48/0.068/ & -1.85/0.138/ & -1.09/0.338/ & -2.36/0.078/ & -6.02/0.004/** & --- \\
\hline
\end{tabular}
}
\caption{\label{tab:cv-female-age}CommonVoice: Pairwise t-tests for female age groups (t / p / sig).}
\end{table}

\begin{table*}[t]
\centering
\resizebox{\textwidth}{!}{
\begin{tabular}{llccccccc}
\hline
\textbf{Gender} & \textbf{Speakers} & \textbf{WavLM-Base} & \textbf{WavLM-Base+} & \textbf{RedimNet} & \textbf{ECAPA} & \textbf{MFA-Conformer} & \textbf{Titanet} \\
\hline
Female & 45 & 28.63\% & 20.36\% & 3.47\% & 5.38\% & 6.85\% & 3.63\% \\
Male   & 21 & 17.42\% & 17.56\% & 5.41\% & 9.84\% & 10.16\% & 9.70\%\\
\hline
\end{tabular}
}
\caption{\label{tab:eer_diff_language}CommonVoice: 
EER when target pairs come from different language while non-target pairs come from the same language.
}
\end{table*}

\begin{table*}[t]
  \centering
  \resizebox{\textwidth}{!}{
  \begin{tabular}{lccccc}
    \hline
    \textbf{TTS System} & \textbf{WavLM-Base} & \textbf{WavLM-Base+} & \textbf{RedimNet} & \textbf{ECAPA-TDNN} & \textbf{MFA-Conformer} \\
    \hline
    CosyTTS & 25.74\% & 23.76\% & 5.94\% & 5.94\% & 0.00\% \\
    StyleTTS    & 23.76\% &	22.77\%	& 3.96\% &	3.96\%	& 0.00\% \\
    xTTS      & 25.74\% &	24.75\% &	5.94\% &	3.96\% &	0.00\% \\
    \hline
  \end{tabular}}
          \caption{\label{tab:tts_conditions}
  EARS: EER of different SV models when evaluated against spoofed speech generated by various TTS systems.
  }
\end{table*}

\begin{table*}[t]
\centering
\resizebox{\textwidth}{!}{
\begin{tabular}{llcccccc}
\hline
\textbf{Conditions} & \textbf{Gender} & \textbf{WavLM-Base} & \textbf{WavLM-Base+} & \textbf{RedimNet} & \textbf{ECAPA} & \textbf{MFA-Conformer} & \textbf{Titanet} \\
\hline
Plain & M & 15.14\% & 16.75\% & 0.26\% & 0.56\% & 0.56\% & 0.60\% \\
Plain & F & 15.28\% & 18.06\% & 0.43\% & 1.05\% & 1.07\% & 0.60\% \\
Lombard & M & 13.16\% & 15.72\% & 0.12\% & 0.57\% & 0.48\% & 0.39\% \\
Lombard & F & 14.11\% & 17.38\% & 0.33\% & 0.99\% & 1.00\% & 0.46\% \\
Mixed & M & 17.30\% & 19.13\% & 0.50\% & 1.68\% & 1.18\% & 1.18\% \\
Mixed & F & 18.27\% & 20.31\% & 0.89\% & 2.50\% & 1.97\% & 1.59\% \\
\hline
\end{tabular}
}
\caption{\label{tab:eer_lombard} 
Lombard Grid: EER in Plain, Lombard and Mixed conditions. Mixed conditions imply that target pairs come from different conditions (one plain, one lombard) while non-target pairs come from the same condition. There are 30 Female and 24 Male speakers in the dataset.
}
\end{table*}

\begin{table*}[t]
\centering
\resizebox{\textwidth}{!}{
\begin{tabular}{llcccccc}
\hline
\textbf{Conditions} & \textbf{Gender} & \textbf{WavLM-Base} & \textbf{WavLM-Base+} & \textbf{RedimNet} & \textbf{ECAPA} & \textbf{MFA-Conformer} & \textbf{Titanet} \\
\hline
Plain & M & 0.9568 & 0.9311 & 0.0421 & 0.0768 & 0.0726 & 0.0676 \\
Plain & F & 0.9652 & 0.9605 & 0.0755 & 0.1473 & 0.1427 & 0.1064 \\
Lombard & M & 0.9317 & 0.9209 & 0.0189 & 0.0840 & 0.0503 & 0.0403 \\
Lombard & F & 0.9369 & 0.9406 & 0.0643 & 0.1157 & 0.1085 & 0.0639 \\
Mixed & M & 0.9824 & 0.9712 & 0.0917 & 0.2362 & 0.1506 & 0.1657 \\
Mixed & F & 0.9934 & 0.9880 & 0.1648 & 0.3080 & 0.2458 & 0.2448 \\
\hline
\end{tabular}
}
\caption{\label{tab:mindcf_lombard} 
Lombard Grid: minDCF in Plain, Lombard and Mixed conditions. Mixed conditions imply that target pairs come from different conditions (one plain, one lombard) while non-target pairs come from the same condition. There are 30 Female and 24 Male speakers in the dataset.
}
\end{table*}

\begin{table*}[t]
\centering
\resizebox{\textwidth}{!}{
\begin{tabular}{llcccccc}
\hline
\textbf{Conditions} & \textbf{Gender} & \textbf{WavLM-Base} & \textbf{WavLM-Base+} & \textbf{RedimNet} & \textbf{ECAPA} & \textbf{MFA-Conformer} & \textbf{Titanet} \\
\hline
Plain & M & 0.9274 & 0.9111 & 1.0000 & 0.9998 & 0.9999 & 0.9998 \\
Plain & F & 0.9248 & 0.8999 & 0.9999 & 0.9995 & 0.9994 & 0.9998 \\
Lombard & M & 0.9421 & 0.9174 & 1.0000 & 0.9998 & 0.9999 & 0.9999 \\
Lombard & F & 0.9337 & 0.9073 & 0.9999 & 0.9995 & 0.9996 & 0.9999 \\
Mixed & M & 0.9087 & 0.8883 & 0.9999 & 0.9987 & 0.9994 & 0.9993 \\
Mixed & F & 0.8981 & 0.8794 & 0.9996 & 0.9972 & 0.9982 & 0.9988 \\
\hline
\end{tabular}
}
\caption{\label{tab:auc_lombard} 
Lombard Grid: ROC AUC in Plain, Lombard and Mixed conditions. Mixed conditions imply that target pairs come from different conditions (one plain, one lombard) while non-target pairs come from the same condition. There are 30 Female and 24 Male speakers in the dataset.
}
\end{table*}

\begin{table*}[t]
\centering
\resizebox{\textwidth}{!}{
\begin{tabular}{llcccccc}
\hline
\textbf{Age}            & \textbf{Gender} & \textbf{WavLM-Base} & \textbf{WavLM-Base+} & \textbf{RedimNet} & \textbf{ECAPA} & \textbf{MFA-Conformer} & \textbf{Titanet} \\
\hline
\multirow{2}{*}{Teens}      & F \hspace{4pt}112 spks    & 0.9996 & 0.9986 & 0.6078 & 0.7261 & 0.7463 & 0.7157 \\
                            & M 112 spks                & 0.9848 & 0.9591 & 0.2597 & 0.3124 & 0.4485 & 0.5234 \\
\multirow{2}{*}{Twenties}   & F \hspace{4pt}582 spks    & 0.9964 & 0.9688 & 0.2987 & 0.4615 & 0.5083 & 0.5726 \\
                            & M 582 spks                & 0.9507 & 0.9140 & 0.2566 & 0.3588 & 0.3650 & 0.4417 \\
\multirow{2}{*}{Thirties}   & F \hspace{4pt}240 spks    & 0.9679 & 0.9239 & 0.2654 & 0.3799 & 0.4101 & 0.4281 \\
                            & M 240 spks                & 0.9543 & 0.9423 & 0.1228 & 0.2232 & 0.2269 & 0.3646 \\
\multirow{2}{*}{Fourties}   & F \hspace{4pt}140 spks    & 0.9377 & 0.9019 & 0.1266 & 0.2420 & 0.2790 & 0.3207 \\
                            & M 140 spks                & 0.9361 & 0.9040 & 0.1766 & 0.3348 & 0.2972 & 0.4120 \\
\multirow{2}{*}{Fifties}    & F \hspace{4pt}110 spks    & 0.9751 & 0.9864 & 0.2604 & 0.4587 & 0.4618 & 0.6016 \\
                            & M 126 spks                & 0.9279 & 0.9175 & 0.1125 & 0.2655 & 0.2333 & 0.4535 \\
\multirow{2}{*}{Sixties}    & F \hspace{9pt}49 spks     & 0.9487 & 0.9095 & 0.3820 & 0.6552 & 0.5506 & 0.8225 \\
                            & M  \hspace{5pt}57 spks    & 0.9840 & 0.9807 & 0.5248 & 0.5615 & 0.5725 & 0.7283 \\
\multirow{2}{*}{Seventy+}   & F  \hspace{9pt}17 spks    & 0.8033 & 0.7541 & 0.3443 & 0.3443 & 0.0492 & 0.1803 \\
                            & M  \hspace{5pt}69 spks    & 0.9645 & 0.9628 & 0.2349 & 0.5360 & 0.3992 & 0.4257 \\
\hline
\end{tabular}
}
\caption{\label{tab:mindcf_age_gender}CommonVoice: 
minDCF variation over age for both genders.
}
\end{table*}

\begin{table*}[t]
\centering
\resizebox{\textwidth}{!}{
\begin{tabular}{llcccccc}
\hline
\textbf{Age}            & \textbf{Gender} & \textbf{WavLM-Base} & \textbf{WavLM-Base+} & \textbf{RedimNet} & \textbf{ECAPA} & \textbf{MFA-Conformer} & \textbf{Titanet} \\
\hline
\multirow{2}{*}{Teens}      & F \hspace{4pt}112 spks    & 0.7526 & 0.7482 & 0.9359 & 0.9340 & 0.9282 & 0.9006 \\
                            & M 112 spks                & 0.8478 & 0.8774 & 0.9836 & 0.9831 & 0.9770 & 0.9537 \\
\multirow{2}{*}{Twenties}   & F \hspace{4pt}582 spks    & 0.8446 & 0.8866 & 0.9922 & 0.9861 & 0.9842 & 0.9421 \\
                            & M 582 spks                & 0.8448 & 0.8638 & 0.9735 & 0.9663 & 0.9693 & 0.9447 \\
\multirow{2}{*}{Thirties}   & F \hspace{4pt}240 spks    & 0.8577 & 0.8625 & 0.9963 & 0.9910 & 0.9863 & 0.9656 \\
                            & M 240 spks                & 0.8785 & 0.9023 & 0.9917 & 0.9889 & 0.9901 & 0.9568 \\
\multirow{2}{*}{Fourties}   & F \hspace{4pt}140 spks    & 0.8711 & 0.8827 & 0.9971 & 0.9889 & 0.9846 & 0.9582 \\
                            & M 140 spks                & 0.9020 & 0.9063 & 0.9982 & 0.9948 & 0.9956 & 0.9737 \\
\multirow{2}{*}{Fifties}    & F \hspace{4pt}110 spks    & 0.8445 & 0.8539 & 0.9968 & 0.9882 & 0.9845 & 0.8915 \\
                            & M 126 spks                & 0.8902 & 0.9129 & 0.9987 & 0.9967 & 0.9935 & 0.9411 \\
\multirow{2}{*}{Sixties}    & F \hspace{9pt}49 spks     & 0.7990 & 0.8069 & 0.9871 & 0.9598 & 0.9579 & 0.8335 \\
                            & M  \hspace{5pt}57 spks    & 0.8810 & 0.9025 & 0.9966 & 0.9913 & 0.9924 & 0.9739 \\
\multirow{2}{*}{Seventy+}   & F  \hspace{9pt}17 spks    & 0.8781 & 0.9218 & 0.9983 & 0.9885 & 0.9987 & 0.9893 \\
                            & M  \hspace{5pt}69 spks    & 0.8449 & 0.8042 & 0.9326 & 0.9247 & 0.9571 & 0.9205 \\
\hline
\end{tabular}
}
\caption{\label{tab:auc_age_gender}CommonVoice: 
ROC AUC variation over age for both genders.
}
\end{table*}

\begin{table*}[t]
\centering
\resizebox{\textwidth}{!}{
\begin{tabular}{llcccccc}
\hline
\textbf{Codec} & \textbf{Condition} & \textbf{WavLM-Base} & \textbf{WavLM-Base+} & \textbf{RedimNet} & \textbf{ECAPA} & \textbf{MFA-Conformer} & \textbf{Titanet}\\
\hline
                        & Clean             & 0.9884 & 0.9549 & 0.2705 & 0.3993 & 0.4258 & 0.2992 \\
 \hline
\multirow{4}{*}{GSM}    & No Noise          & 0.9884 & 0.9551 & 0.2692 & 0.3999 & 0.4282 & 0.2995 \\
                        & GaussNoise+RIR    & 1.0000 & 1.0000 & 0.9941 & 1.0000 & 1.0000 & 1.0000 \\
                        & EnvNoise+RIR      & 1.0000 & 1.0000 & 0.7685 & 0.8536 & 0.9835 & 0.9811 \\
                        & CrossTalk+RIR     & 1.0000 & 1.0000 & 0.8182 & 0.8999 & 0.9893 & 0.9918 \\
                        \hline
\multirow{4}{*}{AMR}    & No Noise          & 0.9884 & 0.9551 & 0.2692 & 0.3999 & 0.4282 & 0.2995 \\
                        & GaussNoise+RIR    & 1.0000 & 1.0000 & 0.9941 & 1.0000 & 1.0000 & 1.0000 \\
                        & EnvNoise+RIR      & 1.0000 & 1.0000 & 0.7685 & 0.8536 & 0.9835 & 0.9811 \\
                        & CrossTalk+RIR     & 1.0000 & 1.0000 & 0.8182 & 0.8999 & 0.9893 & 0.9918 \\
                        \hline
\multirow{4}{*}{Opus}   & No Noise          & 0.9884 & 0.9551 & 0.2692 & 0.3999 & 0.4282 & 0.2995 \\
                        & GaussNoise+RIR    & 1.0000 & 1.0000 & 0.9941 & 1.0000 & 1.0000 & 1.0000 \\
                        & EnvNoise+RIR      & 1.0000 & 1.0000 & 0.7685 & 0.8536 & 0.9835 & 0.9811 \\
                        & CrossTalk+RIR     & 1.0000 & 1.0000 & 0.8182 & 0.8999 & 0.9893 & 0.9918 \\
                        \hline
\hline
\end{tabular}
}
\caption{\label{tab:dcf_codec} CommonVoice: minDCF under audio degradation from codecs and noise conditions.
}
\end{table*}

\begin{table*}[t]
\centering
\resizebox{\textwidth}{!}{
\begin{tabular}{llcccccc}
\hline
\textbf{Codec} & \textbf{Condition} & \textbf{WavLM-Base} & \textbf{WavLM-Base+} & \textbf{RedimNet} & \textbf{ECAPA} & \textbf{MFA-Conformer} & \textbf{Titanet}\\
\hline
                        & Clean             & 0.8461 & 0.8730 & 0.9846 & 0.9794 & 0.9787 & 0.9855 \\
 \hline
\multirow{4}{*}{GSM}    & No Noise          & 0.8458 & 0.8729 & 0.9846 & 0.9793 & 0.9783 & 0.9854 \\
                        & GaussNoise+RIR    & 0.6287 & 0.6816 & 0.8540 & 0.8973 & 0.7955 & 0.8117 \\
                        & EnvNoise+RIR      & 0.6340 & 0.6583 & 0.9000 & 0.9079 & 0.8780 & 0.8524 \\
                        & CrossTalk+RIR     & 0.6323 & 0.6621 & 0.8318 & 0.8319 & 0.8269 & 0.8380 \\
                        \hline
\multirow{4}{*}{AMR}    & No Noise          & 0.8458 & 0.8729 & 0.9846 & 0.9793 & 0.9783 & 0.9854 \\
                        & GaussNoise+RIR    & 0.6287 & 0.6816 & 0.8540 & 0.8973 & 0.7955 & 0.8117 \\
                        & EnvNoise+RIR      & 0.6340 & 0.6583 & 0.9000 & 0.9079 & 0.8780 & 0.8524 \\
                        & CrossTalk+RIR     & 0.6323 & 0.6621 & 0.8318 & 0.8319 & 0.8269 & 0.8380 \\
                        \hline
\multirow{4}{*}{Opus}   & No Noise          & 0.8458 & 0.8729 & 0.9846 & 0.9793 & 0.9783 & 0.9854 \\
                        & GaussNoise+RIR    & 0.6287 & 0.6816 & 0.8540 & 0.8973 & 0.7955 & 0.8117 \\
                        & EnvNoise+RIR      & 0.6340 & 0.6583 & 0.9000 & 0.9079 & 0.8780 & 0.8524 \\
                        & CrossTalk+RIR     & 0.6323 & 0.6621 & 0.8318 & 0.8319 & 0.8269 & 0.8380 \\
                        \hline
\hline
\end{tabular}
}
\caption{\label{tab:auc_codec} CommonVoice: ROC AUC under audio degradation from codecs and noise conditions.
}
\end{table*}

\begin{table*}[t]
\centering
\resizebox{\textwidth}{!}{
\begin{tabular}{llccccccc}
\hline
\textbf{Noise} & \textbf{SNR} & \textbf{RIR} & \textbf{WavLM-Base} & \textbf{WavLM-Base+} & \textbf{RedimNet} & \textbf{ECAPA} & \textbf{MFA-Conformer} & \textbf{Titanet} \\
\hline
Clean                       &    &   & 23.05\% & 20.23\% &  4.69\% &  6.13\% &  6.65\% & 4.92\% \\
\hline
\multirow{3}{*}{GaussNoise} &  5 &   & 31.86\% & 34.53\% & 21.75\% & 18.95\% & 31.51\% & 17.35\% \\
                            & 15 &   & 32.00\% & 31.66\% & 16.02\% & 12.66\% & 18.85\% & 11.92\% \\
                            & 25 &   & 30.11\% & 29.75\% &  9.47\% &  9.21\% & 12.03\% & 7.98\% \\
\hline
\multirow{3}{*}{GaussNoise w/ RIR} &  5 & 2 & 46.57\% & 47.67\% & 28.19\% & 21.89\% & 35.36\% & 44.05\% \\
                            & 15 & 3 & 40.86\% & 36.91\% & 22.52\% & 18.03\% & 27.54\% & 26.54\% \\
                            & 25 & 4 & 46.42\% & 44.18\% & 42.82\% & 35.50\% & 37.37\% & 43.09\% \\
\hline
\multirow{3}{*}{EnvNoise}   &  5 &   & 42.01\% & 43.05\% & 25.04\% & 24.46\% & 26.70\% & 23.56\% \\
                            & 15 &   & 34.79\% & 32.34\% & 10.52\% & 13.62\% & 15.34\% & 11.19\% \\
                            & 25 &   & 27.30\% & 23.15\% &  5.95\% &  8.49\% &  9.59\% & 6.32\% \\
\hline
\multirow{3}{*}{EnvNoise w/ RIR} &  5 & 2 & 45.17\% & 46.32\% & 32.75\% & 25.82\% & 27.26\% & 36.22\% \\
                            & 15 & 3 & 40.25\% & 38.65\% & 19.36\% & 15.88\% & 20.04\% & 22.52\% \\
                            & 25 & 4 & 46.37\% & 44.19\% & 39.43\% & 35.93\% & 36.71\% & 43.14\% \\
\hline
\multirow{3}{*}{CrossTalk}  &  5 &   & 47.54\% & 47.35\% & 38.97\% & 37.97\% & 36.15\% & 36.97\% \\
                            & 15 &   & 43.41\% & 41.61\% & 20.50\% & 26.00\% & 24.11\% & 22.67\% \\
                            & 25 &   & 35.37\% & 29.27\% &  8.59\% & 15.06\% & 14.02\% & 10.06\% \\
\hline
\multirow{3}{*}{CrossTalk w/ RIR} &  5 & 2 & 46.19\% & 46.75\% & 41.23\% & 37.73\% & 36.13\% & 38.73\% \\
                            & 15 & 3 & 40.54\% & 38.48\% & 26.19\% & 24.54\% & 24.63\% & 23.71\% \\
                            & 25 & 4 & 46.44\% & 44.20\% & 39.25\% & 36.55\% & 37.41\% & 43.92\% \\ \hline

\end{tabular}
}
\caption{\label{tab:eer_noise}CommonVoice: 
EER degradation due to noise and room reverberations.
}
\end{table*}

\begin{table*}[t]
\centering
\resizebox{\textwidth}{!}{
\begin{tabular}{llccccccc}
\hline
\textbf{Noise} & \textbf{SNR} & \textbf{RIR} & \textbf{WavLM-Base} & \textbf{WavLM-Base+} & \textbf{RedimNet} & \textbf{ECAPA} & \textbf{MFA-Conformer} & \textbf{Titanet} \\
\hline
Clean                       &    &          & 0.9884 & 0.9549 & 0.2705 & 0.3993 & 0.4258 & 0.2992 \\
\hline
\multirow{3}{*}{GaussNoise} &  5 &          & 1.0000 & 1.0000 & 1.0000 & 1.0000 & 1.0000 & 1.0000 \\
                            & 15 &          & 1.0000 & 1.0000 & 0.8279 & 1.0000 & 1.0000 & 0.7479 \\
                            & 25 &          & 0.9997 & 0.9955 & 0.5782 & 0.6777 & 0.8822 & 0.5019 \\
\hline
\multirow{3}{*}{GaussNoise w/ RIR} &  5 & 2 & 1.0000 & 1.0000 & 1.0000 & 1.0000 & 1.0000 & 1.0000 \\
                            & 15 & 3        & 1.0000 & 1.0000 & 0.9968 & 1.0000 & 1.0000 & 1.0000 \\
                            & 25 & 4        & 1.0000 & 1.0000 & 1.0000 & 1.0000 & 1.0000 & 1.0000 \\
\hline
\multirow{3}{*}{EnvNoise}   &  5 &          & 1.0000 & 1.0000 & 0.8034 & 0.9714 & 1.0000 & 0.8525 \\
                            & 15 &          & 1.0000 & 0.9901 & 0.4769 & 0.6297 & 0.7184 & 0.5016 \\
                            & 25 &          & 0.9968 & 0.9705 & 0.3288 & 0.4782 & 0.5360 & 0.3574 \\
\hline
\multirow{3}{*}{EnvNoise w/ RIR} &  5 & 2   & 1.0000 & 1.0000 & 1.0000 & 1.0000 & 1.0000 & 1.0000 \\
                            & 15 & 3        & 1.0000 & 1.0000 & 0.8247 & 0.8473 & 0.9861 & 0.9809 \\
                            & 25 & 4        & 1.0000 & 1.0000 & 1.0000 & 1.0000 & 1.0000 & 1.0000 \\
\hline
\multirow{3}{*}{CrossTalk}  &  5 &          & 1.0000 & 1.0000 & 0.9043 & 1.0000 & 1.0000 & 1.0000 \\
                            & 15 &          & 1.0000 & 0.9989 & 0.5424 & 0.8131 & 0.8315 & 0.7161 \\
                            & 25 &          & 0.9978 & 0.9759 & 0.3430 & 0.5748 & 0.6117 & 0.4328 \\
\hline
\multirow{3}{*}{CrossTalk w/ RIR} &  5 & 2  & 1.0000 & 1.0000 & 1.0000 & 1.0000 & 1.0000 & 1.0000 \\
                            & 15 & 3        & 1.0000 & 1.0000 & 0.8752 & 0.8970 & 0.9843 & 0.9893 \\
                            & 25 & 4        & 1.0000 & 1.0000 & 1.0000 & 1.0000 & 1.0000 & 1.0000 \\ \hline

\end{tabular}
}
\caption{\label{tab:dcf_noise}CommonVoice: 
minDCF degradation due to noise and room reverberations.
}
\end{table*}

\begin{table*}[t]
\centering
\resizebox{\textwidth}{!}{
\begin{tabular}{llccccccc}
\hline
\textbf{Noise} & \textbf{SNR} & \textbf{RIR} & \textbf{WavLM-Base} & \textbf{WavLM-Base+} & \textbf{RedimNet} & \textbf{ECAPA} & \textbf{MFA-Conformer} & \textbf{Titanet} \\
\hline
Clean                       &    &          & 0.8461 & 0.8730 & 0.9846 & 0.9794 & 0.9787 & 0.9855 \\
\hline
\multirow{3}{*}{GaussNoise} &  5 &          & 0.7448 & 0.7166 & 0.8595 & 0.8765 & 0.7493 & 0.8965 \\
                            & 15 &          & 0.7427 & 0.7507 & 0.9162 & 0.9415 & 0.8855 & 0.9485 \\
                            & 25 &          & 0.7684 & 0.7760 & 0.9634 & 0.9654 & 0.9475 & 0.9723 \\
\hline
\multirow{3}{*}{GaussNoise w/ RIR} &  5 & 2 & 0.5483 & 0.5395 & 0.7872 & 0.8490 & 0.7043 & 0.5977 \\
                            & 15 & 3        & 0.6253 & 0.6783 & 0.8535 & 0.8961 & 0.7927 & 0.8073 \\
                            & 25 & 4        & 0.5576 & 0.5773 & 0.6039 & 0.6746 & 0.6425 & 0.6039 \\
\hline
\multirow{3}{*}{EnvNoise}   &  5 &          & 0.6140 & 0.6115 & 0.8228 & 0.8350 & 0.8103 & 0.8419 \\
                            & 15 &          & 0.6992 & 0.7297 & 0.9432 & 0.9280 & 0.9184 & 0.9454 \\
                            & 25 &          & 0.7911 & 0.8369 & 0.9761 & 0.9625 & 0.9609 & 0.9771 \\
\hline
\multirow{3}{*}{EnvNoise w/ RIR} &  5 & 2   & 0.5693 & 0.5551 & 0.7407 & 0.8192 & 0.8016 & 0.6823 \\
                            & 15 & 3        & 0.6343 & 0.6581 & 0.8775 & 0.9124 & 0.8787 & 0.8523 \\
                            & 25 & 4        & 0.5572 & 0.5709 & 0.6252 & 0.6738 & 0.6620 & 0.6027 \\
\hline
\multirow{3}{*}{CrossTalk}  &  5 &          & 0.5392 & 0.5442 & 0.6778 & 0.6788 & 0.6996 & 0.6941 \\
                            & 15 &          & 0.6018 & 0.6307 & 0.8686 & 0.8210 & 0.8416 & 0.8562 \\
                            & 25 &          & 0.6965 & 0.7665 & 0.9589 & 0.9186 & 0.9301 & 0.9563 \\
\hline
\multirow{3}{*}{CrossTalk w/ RIR} &  5 & 2  & 0.5544 & 0.5490 & 0.6329 & 0.6809 & 0.6987 & 0.6526 \\
                            & 15 & 3        & 0.6318 & 0.6601 & 0.8158 & 0.8349 & 0.8319 & 0.8426 \\
                            & 25 & 4        & 0.5567 & 0.5728 & 0.6244 & 0.6677 & 0.6606 & 0.5965 \\ \hline

\end{tabular}
}
\caption{\label{tab:auc_noise}CommonVoice: 
ROC AUC degradation due to noise and room reverberations.
}
\end{table*}

\subsection{Architectures}
\label{sec:appendix-arch}
We evaluate a diverse set of state-of-the-art SV models that represent different architectural paradigms and training strategies. This choice ensures that our robustness analysis captures not only performance differences, but also the ways in which model design and data exposure influence vulnerability to various stressors.

\textbf{WavLM-Base / WavLM-Base+} are large-scale self-supervised models trained on 94k hours of speech. Their transformer-based encoders achieve strong performance in clean conditions, but because training emphasizes large-scale coverage rather than targeted augmentation, we observe sharp degradations under codec compression and environmental noise. This suggests that pretraining alone is insufficient for robustness when deployment scenarios differ substantially from pretraining corpora.

\textbf{ECAPA-TDNN} incorporates channel attention and multi-layer aggregation, explicitly designed to emphasize salient spectral-temporal cues. As expected, this architectural bias provides stronger resilience under reverberation and far-field settings. However, its convolutional design without adversarial exposure leaves it vulnerable to black-box spoofing (e.g., FakeBob), where we observe the highest error rates among all models.

\textbf{RedimNet} leverages reshape-dimension strategies combined with noise-augmented training data. This enables strong robustness in noisy and codec-mismatched settings, consistent with its training objectives. At the same time, its limited adversarial training leaves it more exposed to white-box perturbations such as FGSM.

\textbf{MFA-Conformer} is trained in-house due to the absence of public checkpoints. Its multi-scale feature aggregation and inclusion of codec/noise-augmented training data provide superior robustness across nearly all stressors. In particular, it achieves zero EER under TTS spoofing (CosyVoice, xTTS, and StyleTTS and the lowest error rates under adversarial attacks, reflecting how targeted training diversity and architectural flexibility together yield state-of-the-art robustness.

\textbf{Titanet} is a hybrid architecture that integrates TDNN layers with attention-based pooling and refined training objectives. Its design is optimized for robustness in speaker embedding extraction, and it shows relatively stable performance across codec and demographic variations. However, Titanet still exhibits elevated error rates in extreme noise + reverberation conditions, indicating that while architectural innovations improve baseline resilience, exposure to diverse training stressors remains critical for full robustness.

\end{document}